\RequirePackage{fix-cm}

\documentclass[smallextended]{svjour3}       
\smartqed  
\usepackage{graphicx}

\usepackage[english]{babel}
\usepackage[utf8]{inputenc}
\usepackage{amsmath}

\usepackage{geometry}
\usepackage[most]{tcolorbox}
\usepackage{fancyhdr}
\usepackage{longtable}
\usepackage{graphicx}
\usepackage{xcolor}
\usepackage{tikz}
\usepackage{pgfgantt}
\usepackage{float}
\usepackage{enumitem}
\usepackage{booktabs}
\usepackage{tabularx}
\usepackage{tabularray}
\usepackage{ragged2e}

\usepackage{natbib}
\usepackage{eurosym}
\usepackage{url}

\usepackage[hidelinks]{hyperref}

\usepackage{listings}
\lstdefinelanguage{json}{
  basicstyle=\ttfamily\small,
  numbers=left,
  numberstyle=\tiny,
  stepnumber=1,
  numbersep=6pt,
  showstringspaces=false,
  breaklines=true,
  frame=single,
  backgroundcolor=\color{gray!5},
  stringstyle=\color{blue},
  keywordstyle=\color{black},
  morekeywords={true,false,null}
}

\begin{document}

\title{Detect--Repair--Verify for LLM-Generated Code: A Multi-Language, Multi-Granularity Empirical Study
}


\author{Cheng Cheng 
}

\institute{Cheng Cheng \at
              Department of Computer Science and \\ Software Engineering, Concordia University,\\Montreal, Canada\\
              \email{cheng.cheng.20171@mail.concordia.ca}           
}
\date{Received: date / Accepted: date}

\maketitle

\begin{abstract}
Large language models can generate runnable software artifacts, but the security of such artifacts is still difficult to evaluate in an end-to-end way. This study investigates that problem through a Detect--Repair--Verify (DRV) workflow, in which vulnerabilities are detected, repaired, and then checked again through security and functional tests. The study addresses four gaps in current evidence: the lack of test-grounded benchmarks for LLM-generated artifacts, especially at broader artifact scales; limited evidence on pipeline-level effectiveness beyond studying detection or repair in isolation; unclear reliability of detection reports as repair guidance; and uncertain repair trustworthiness under verification.
To support this study, the EduCollab benchmark is constructed as a multi-language, multi-granularity benchmark of runnable LLM-generated web-application artifacts in PHP, JavaScript, and Python. Each artifact is paired with executable functional and exploit test suites, and the benchmark spans project-level, requirement-level, and file-level settings. On this benchmark, the study compares unrepaired baselines, single-pass detect--repair, and bounded iterative DRV under comparable budget constraints. Outcomes are measured using secure-and-correct yield, and intermediate artifacts and iteration traces are analyzed to assess report actionability and repair failure modes.
The results show that bounded iterative DRV can improve secure-and-correct yield over single-pass repair, but the improvement is uneven at the project level and becomes clearer at narrower repair scopes. Detection reports are often useful for downstream repair, but their reliability is less consistent, and even detailed reports do not always lead to secure-and-correct outcomes. Repair trustworthiness also depends strongly on repair scope, with the weakest outcomes at the project level and the strongest at the file level. These findings highlight the need for test-grounded, end-to-end evaluation of LLM-based vulnerability management workflows.
\end{abstract}

\section{Introduction}
\label{intro}

Large language models are now part of everyday software development. They can translate natural-language requirements into runnable code and speed up prototyping and implementation across a wide range of tasks, from writing individual functions to assembling complete, project-level artifacts. In this setting, security should be understood through the lens of vulnerability management, which treats software protection as a continuous process rather than a single step at coding time. A typical vulnerability management workflow starts with identifying security-relevant issues during development and maintenance, then detecting potential vulnerabilities in code or running systems, analyzing and localizing the root cause to specific components or code regions, and prioritizing what to handle first based on risk and impact. It then moves to remediation, where developers repair vulnerabilities by applying patches or refactoring insecure logic, followed by verification to ensure the fix is effective and does not break intended behavior. Finally, the workflow includes fix identification and tracking, which links patches to vulnerabilities for auditing, release management, and downstream propagation, so that related systems and versions can be updated consistently. When LLMs participate in software creation and repair, they enter this same lifecycle, and securing LLM-generated code, which is operationalized here as edit-level generation through patches, requires the same end-to-end management across detection, repair, and verification.

Building on this lifecycle perspective, vulnerability management for LLM-generated project artifacts can be operationalized as a structured Detect–Repair–Verify (DRV) workflow. Starting from an initial artifact, whether generated or manually implemented, potential vulnerabilities are identified through automated tools or model-based analysis. The resulting reports then guide targeted repair actions, which may also be assisted by LLMs. Finally, the repaired artifact is re-evaluated through security checks and functional tests to confirm that vulnerabilities are mitigated and intended behavior is preserved. This Detect–Repair–Verify workflow reflects how vulnerability management is enacted in practice when LLM-generated artifacts enter development pipelines, and it provides a concrete basis for multi-language empirical evaluation under comparable verification protocols. In practice, this workflow may begin under two common conditions. Vulnerabilities may remain latent when an artifact is produced, in which case detection is needed before repair can proceed. Alternatively, vulnerabilities may already be known through prior analysis, security testing, or exploit confirmation, shifting the focus from discovery to whether repair is effective and trustworthy under verification.

Despite the rapid uptake of LLMs in programming workflows, existing studies still provide limited evidence on the effectiveness of end-to-end security-hardening pipelines for LLM-generated artifacts. Much of the literature examines vulnerability detection and vulnerability repair as separate tasks, rather than as a composed generate--detect--repair--verify workflow of the kind used in practice. This limitation matters because detection outputs can be unstable and sensitive to superficial code variations, reducing their reliability as repair guidance and creating risks of missed vulnerabilities or spurious alarms \citep{ullah2024sp_secLLMHolmes}. At the same time, repair-oriented studies increasingly emphasize post-repair validation, such as rerunning security checks and functional tests, to confirm mitigation and rule out regressions or unintended side effects \citep{wang2025naacl_cvebench,kim2025usenix_san2patch,zhang2024nist_vulrepair,weI2025arxiv_patcheval}. Taken together, these observations motivate evaluating the workflow as an integrated process, with attention to pipeline-level effectiveness, the reliability of detection reports as repair guidance, and the verified quality of repaired outputs.

\textbf{Limitation 1 (Lack of test-grounded benchmarks, especially at broader artifact scales).}
Most existing evaluations rely on simplified artifacts (e.g., isolated functions or snippets) or benchmarks without executable, security-relevant test oracles. This limits the ability to study LLM-generated artifacts under realistic verification conditions, where functional tests and security tests jointly determine whether an artifact is both correct and secure. Although this limitation is especially pronounced at the project level, benchmark support remains limited more broadly across different artifact granularities.

\textbf{Limitation 2 (Unverified pipeline-level effectiveness across repair scopes).}
Existing studies typically assess vulnerability detection or vulnerability repair in isolation. Consequently, evidence is still limited on whether an end-to-end generate--detect--repair--verify workflow consistently improves the secure-and-correct yield of LLM-generated artifacts over generation-only baselines, and how such effects vary across different repair scopes.

\textbf{Limitation 3 (Unclear detection reliability as repair guidance across granularities).}
Detection reports on LLM-generated code can be unstable and uneven in evidential quality. When used as repair guidance, missed findings may leave vulnerabilities unaddressed, while spurious findings can prompt unnecessary or misdirected edits. It also remains unclear how reliably such reports support downstream repair when the available context and repair target vary across project-, requirement-, and file-level settings.

\textbf{Limitation 4 (Uncertain repair trustworthiness and side effects under verification).}
LLM-based repair may remove reported issues, but it can also introduce regressions, semantic drift, or new security flaws. As a result, the trustworthiness of repaired artifacts under verification, and the failure modes that dominate in practice, remain insufficiently characterized, especially when repair is performed under different context scopes and locality constraints.

These limitations motivate a systematic empirical study of LLM-based detect-and-repair workflows as an integrated process, focusing on when they improve security and correctness, how detection quality affects downstream repair, and why repair succeeds or fails across different artifact granularities.
Accordingly, the study is structured around the following research questions:

\noindent\textbf{RQ1 (Pipeline-level effectiveness).}
Under comparable budget constraints, does bounded iterative detect--repair--verify with test-grounded feedback improve the secure-and-correct yield of LLM-generated artifacts over a single-pass detect--repair baseline, and how does this effect vary across prompt granularities?

\noindent\textbf{RQ2 (Detection reliability as repair guidance).}
How reliable and actionable are LLM-generated detection reports as guidance for downstream repair across different artifact granularities?

\noindent\textbf{RQ3 (Repair trustworthiness under verification).}
To what extent can LLM-based repair mitigate reported vulnerabilities in iterative workflows without introducing functional regressions, and what failure patterns are observed under verification?

To answer these questions, the study evaluates detect--repair--verify in a bounded iterative setting, where the maximum number of iterations is fixed to keep the comparison fair across settings. The evaluation includes an unrepaired baseline and several iterative DRV settings with different iteration budgets, so that single-pass repair can be treated as the one-iteration case of the same framework. To make this evaluation possible, the study constructs the EduCollab benchmark, which contains runnable LLM-generated web-application artifacts together with executable functional and exploit test suites. The benchmark covers PHP, JavaScript, and Python, and includes project-level, requirement-level, and file-level settings so that the study can compare different amounts of context and different repair scopes.

The results show a clear but uneven pattern. For RQ1, bounded iterative detect--repair--verify improves secure-and-correct yield over single-pass repair in some settings, but the benefit is not uniform at the project level and becomes clearer as the repair target becomes smaller, moving from projects to requirements and files. For RQ2, detection reports are often useful for downstream repair because they usually provide explicit fault localization and repair hints, but their reliability is less consistent, and even detailed reports do not always lead to secure-and-correct outcomes under verification. For RQ3, LLM-based repair can produce trustworthy fixes, but the outcome depends strongly on repair scope. Trustworthiness is lowest at the project level, improves at the requirement level, and is highest at the file level. The main remaining difficulties come from cases that do not reach a secure-and-correct state within the bounded budget, and from cases that need additional iterations before both security and functionality are restored.

The remainder of this paper is organized as follows. Section~\ref{related} reviews prior work on LLM-assisted vulnerability detection, automated repair, and verification-oriented evaluation. Section~\ref{motivation} presents motivating scenarios for LLM-generated code security. Section~\ref{methodology} describes the experimental design, including the studied workflows and baselines, budget and iteration constraints, prompt granularities, subject systems, and the functional and security test suites used for verification. Section~\ref{Results} reports the empirical results for RQ1--RQ3, covering pipeline-level secure-and-correct yield, detection reliability as repair guidance, and repair trustworthiness with dominant failure modes. Section~\ref{disscusion} discusses the main findings and their implications for LLM-based vulnerability management workflows. Section~\ref{threats} discusses threats to validity and mitigating measures. Section~\ref{conclusion} concludes with key findings and implications for LLM-based vulnerability management workflows. The benchmark and accompanying artifacts are available in a public repository.\footnote{\url{https://github.com/Hahappyppy2024/EmpricalVDR}}

\section{Related Work}
\label{related}

Large language models (LLMs) are increasingly used in security-relevant software engineering tasks. Existing work spans four closely related directions: secure code generation, vulnerability detection and localization, automated vulnerability repair, and prompting or workflow strategies that steer model behavior. This study connects these directions by examining detect--repair--verify workflows with test-grounded evidence across multiple language implementations and prompt granularities. The rest of this section reviews prior work in each area and clarifies how it relates to the proposed evaluation setting.

\subsection{Secure Code Generation by LLMs}
A growing body of work investigates whether LLM-generated code is secure by default and how to improve its security properties. The literature increasingly moves beyond functional correctness to evaluation protocols that jointly assess correctness and security using testable evidence. SecurityEval provides CWE-aligned prompts and examples to probe vulnerability-prone generations, while SALLM further systematizes the evaluation pipeline and metrics to reduce reliance on ad hoc rules or subjective judgments~\citep{siddiq2022securityeval,siddiq2024sallm}. Along this evaluation-centric line, Cheng et al. propose benchmarking security-relevant behaviors in LLM-based code generation~\citep{cheng2024benchmarking}, and CFCEval offers a more structured set of security-oriented criteria with evidence-based checks for assessing whether generated implementations satisfy security expectations~\citep{cheng2025cfceval}. Beyond benchmarks, recent efforts incorporate security into model training and generation-time control: SafeCoder improves security via instruction tuning~\citep{he2024safecoder}, whereas SCodeGen introduces constrained decoding to steer outputs toward trustworthy implementations in real time~\citep{SCodeGen2025Qu}. Finally, newer benchmarks such as BaxBench and CWEval push toward more realistic and reproducible assessment by combining executable tests with exploit-oriented or outcome-driven verification, highlighting the prevalence of code that is correct yet insecure~\citep{vero2025baxbench,peng2025cweval}. Complementing these benchmark-driven studies, empirical evidence from S\&P/CCS shows that AI coding assistants measurably affect developers' security behavior and the security quality of their outputs, suggesting that security cannot be assumed by default~\citep{pearce2022copilot,perry2023insecureusers}. Collectively, these findings motivate security-aware generation and rigorous, test-grounded evaluation, aligning with the need for evidence-based workflows when studying detect--repair--verify pipelines.

\subsection{Vulnerability Detection}

Prior work on vulnerability detection can be broadly categorized into traditional static analysis, dynamic analysis, learning-based detection, and recent LLM-based approaches, which differ in their reliance on program semantics, runtime behavior, and data-driven modeling.

\paragraph{Traditional static analysis.}
Static vulnerability detection analyzes program code without execution and relies on explicit program semantics such as data flow and taint propagation. STASE~\citep{stase2024ase} exemplifies this line by applying static taint analysis to identify configuration-related vulnerabilities, offering interpretable and scalable detection while remaining sensitive to analysis precision and contextual complexity.

\paragraph{Dynamic analysis.}
Dynamic approaches identify vulnerabilities by observing runtime behavior. DeFiWarder~\citep{defiwarder2023ase} illustrates this paradigm by detecting token-level vulnerabilities in DeFi applications through runtime semantics, though dynamic techniques are inherently limited by execution coverage and input exploration.

\paragraph{Learning-based detection.}
Learning-based methods infer vulnerability patterns from data using learned representations of programs, commonly built on structured graphs. MVD leverages flow-sensitive graph neural networks for memory vulnerability detection \citep{mvd2022icse}. Later work adds explanation mechanisms to improve interpretability and support analyst-facing use, as exemplified by Coca and CFExplainer \citep{coca2024icse,cfexplainer2024acm}. In the context of detect--repair workflows, an additional consideration is the quality of detection outputs as actionable guidance, including the stability of findings, the specificity of evidence, and the precision of localization.

\paragraph{Vulnerability Localization}
Vulnerability localization identifies the precise code locations responsible for security vulnerabilities—typically at the function-, statement-, or line-level—and serves as a critical bridge between vulnerability detection and subsequent remediation. Compared with coarse-grained detection, localization places stronger requirements on semantic understanding and granularity, since it must not only determine whether a vulnerability exists but also pinpoint where it manifests in the program. Early learning-based approaches model vulnerable code patterns directly for fine-grained localization: VulDeeLocator performs statement-level localization with neural models over code representations, showing that learned features can move beyond file- or function-level detection~\citep{li2022vuldeelocator}. Building on this direction, VulTeller couples localization with vulnerability description generation, enabling models to both highlight vulnerable regions and produce natural-language explanations of the underlying issues~\citep{zhang2023vulteller}. More recent studies explore large language models to improve localization by leveraging stronger contextual reasoning and configuration choices: ENVUL examines LLM-based localization and studies how task-specific tuning interacts with prompting strategies to yield effective setups~\citep{tian2025envul}. In parallel, VulPCL jointly tackles vulnerability prediction, categorization, and localization, integrating localization with complementary analysis tasks to provide more actionable code-level outputs~\citep{liu2024vulpcl}. Overall, the literature trends toward more context-aware and holistic localization methods that connect detection signals to actionable remediation guidance.

\subsection{Vulnerability Repair}

Automated vulnerability repair aims to generate or recommend code changes that eliminate security vulnerabilities while preserving program functionality. Early research in this area spans a wide range of techniques, including rule-based patch generation, pattern mining, and learning-based approaches. A comprehensive overview of this landscape is provided by the SoK~\citep{Hu0SGZXY025} on automated vulnerability repair, which systematically reviews existing methods, tools, benchmarks, and evaluation practices, and highlights common challenges such as patch correctness, generalization, and evaluation reliability.

\subsubsection{Pattern- and learning-based vulnerability repair.}
Learning-based repair approaches have gained attention due to their ability to generalize repair patterns from data. VulRepair~\citep{fu2022vulrepair} represents a representative sequence-to-sequence approach that formulates vulnerability repair as a code translation problem using a T5-based model, demonstrating the feasibility of neural models for generating vulnerability-fixing patches. Complementary to direct patch generation, VulMatch~\citep{cao2025vulmatch} focuses on extracting and matching repair patterns from historical fixes, leveraging recurring fix templates to enhance repair effectiveness and interpretability.

\subsubsection{LLM-based Vulnerability Repair}
\paragraph{Early exploration of LLM capabilities for repair.}
In parallel, early studies have explored the repair capabilities of large language models without task-specific fine-tuning. Pearce et al.~\citeyearpar{pearce2021zeroshot} examine zero-shot vulnerability repair with large language models, establishing an important baseline and revealing both the potential and limitations of LLMs when applied to vulnerability repair without explicit training signals. Together, these works illustrate the evolution of automated vulnerability repair from pattern- and model-driven approaches toward more flexible, data-driven repair paradigms.

\paragraph{Empirical understanding of LLM-based repair.}
More recent work explicitly investigates large language models as a central component in automated vulnerability repair. VRpilot~\citep{kulsum2024vrpilot} presents an in-depth case study of LLM-based vulnerability repair, analyzing how reasoning strategies and patch validation feedback influence repair performance and reliability. Rather than proposing a single end-to-end solution, this work provides empirical insights into the behavior of LLMs in vulnerability repair settings.

\paragraph{Prompt-driven and iterative LLM repair.}
Subsequent studies aim to enhance LLM-based repair through improved prompting and adaptation strategies. De-Fitero-Dominguez et al.~\citeyearpar{defiterodominguez2024enhancedavr} propose an approach that leverages large language models for automated code vulnerability repair, demonstrating that careful prompt design and task formulation can substantially affect repair outcomes. LLM4CVE~\citep{fakih2025llm4cve} further extends this direction by enabling iterative vulnerability repair, where LLMs interact with vulnerability information and intermediate feedback to progressively refine repair candidates.

\paragraph{CWE-guided and fine-grained control of LLM repair.}
Beyond prompting and iteration, recent work explores more fine-grained control of LLM behavior for vulnerability repair. The fine-grained cue optimization approach combines localized cues with sensitive fine-tuning strategies to guide LLMs toward more accurate and targeted repairs, aiming to improve patch correctness and stability across different vulnerability types
~\citep{zhang2025cueopt}. 
Overall, these studies indicate a shift from one-shot patch generation toward iterative, feedback-driven, and cue-aware LLM-based vulnerability repair frameworks.

\subsection{Prompting Engineering}

Early studies on prompting established its effectiveness as an alternative to fine-tuning for controlling model behavior. A comprehensive overview of prompt-based learning is provided by Liu et al.~\citeyearpar{liu2023pretrainpromptpredict}, who systematically survey prompting methods and categorize them according to their design principles and application scenarios.

A major line of work focuses on eliciting multi-step reasoning through structured prompts. Chain-of-thought~ \citep{wei2022cot} prompting demonstrates that explicitly prompting intermediate reasoning steps can significantly improve performance on reasoning-intensive tasks. Subsequent work shows that large language models can perform such reasoning even in zero-shot settings by using simple trigger phrases \citep{kojima2022zeroshotcot}. To improve the robustness of reasoning, self-consistency aggregates multiple reasoning paths and selects consistent outcomes \citep{wang2023selfconsistency}, while least-to-most prompting decomposes complex problems into simpler subproblems to enable progressive reasoning \citep{zhou2023leasttomost}. Tree-of-thoughts~\citep{yao2023tot} further generalizes these ideas by framing reasoning as a structured search over multiple intermediate thought trajectories.

Beyond reasoning elicitation, recent studies explore prompt optimization and reuse. Wan et al.~\citeyearpar{wan2024teachbetter} investigate the relative roles of instructions and exemplars in automatic prompt optimization, while localized zeroth-order prompt optimization proposes black-box optimization strategies for improving prompts without gradient access \citep{hu2024zopo}. Buffer-of-thoughts~\citep{yang2024bot} introduces reusable reasoning templates to enhance thought-augmented reasoning across tasks , and dynamic rewarding with prompt optimization enables tuning-free self-alignment by adjusting prompts based on reward feedback \citep{singla2024drpo}. Complementary to these approaches, DSPy~\citep{khattab2024dspy} treats prompting as a programmable and optimizable pipeline, compiling declarative specifications into self-improving prompt workflows .

\section{Motivating Scenarios for LLM-Generated Code Security}
\label{motivation}

\begin{figure}
    \centering
    \includegraphics[width=\linewidth]{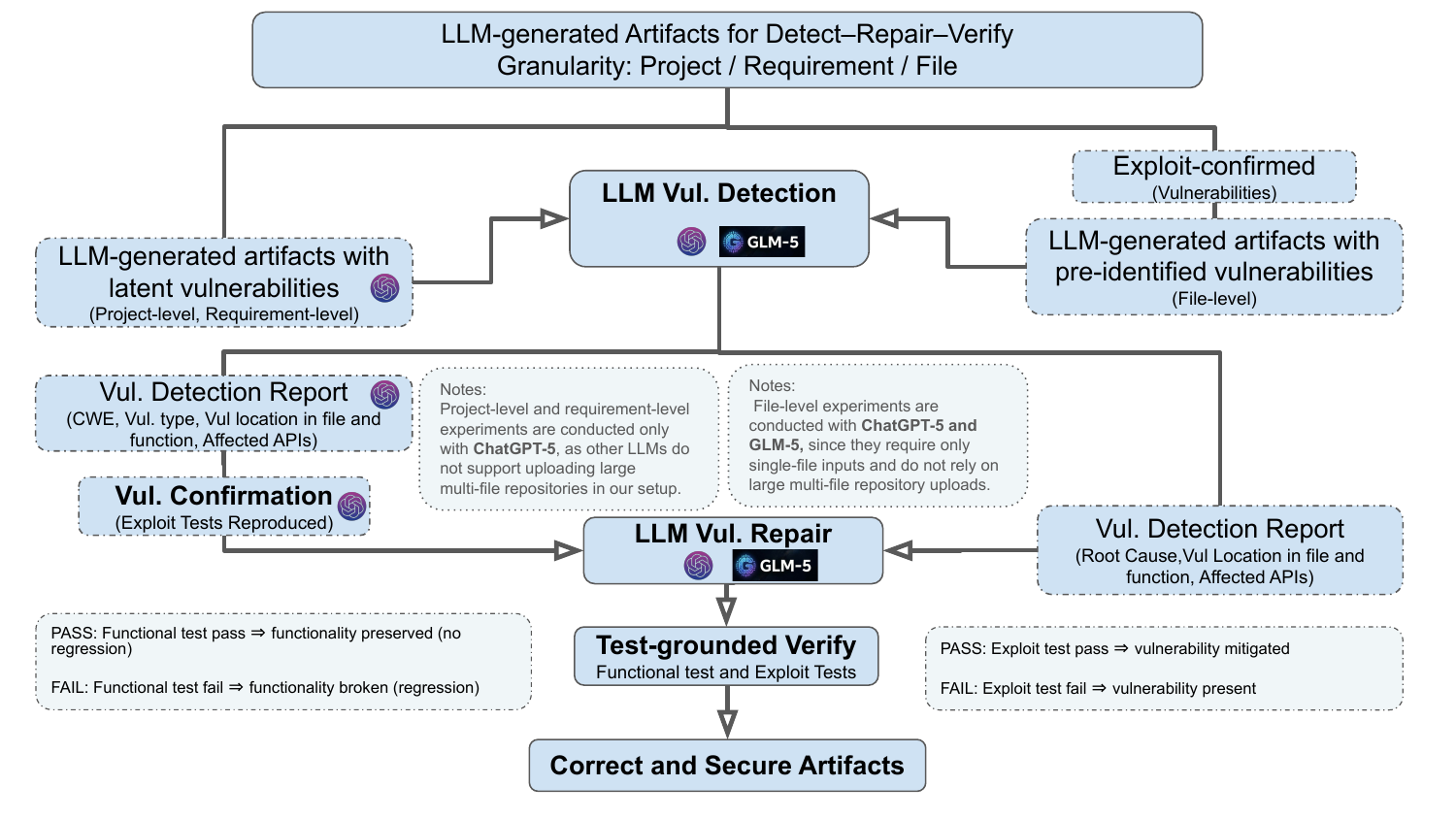}
    \caption{Two motivating scenarios for securing LLM-generated artifacts through detect--repair--verify. In one scenario, vulnerabilities remain latent and must first be identified before repair can begin. In the other, vulnerabilities are already known, so repair starts from explicit guidance. Although the entry points differ, both scenarios are handled within the same detect--repair--verify framework and are evaluated against the same outcome: whether the patched artifact preserves intended functionality while mitigating the targeted vulnerability.}
    \label{fig:motivating_scenarios}
\end{figure}

LLM-generated artifacts may be runnable and feature-complete, yet still contain exploitable vulnerabilities. Security therefore remains a post-generation concern. After an artifact is produced, the question is not only whether it satisfies functional requirements, but also whether vulnerabilities can be identified, repaired, and verified without disrupting intended behavior. This motivates studying LLM-generated code security through a detect–repair–verify perspective.

Two situations are particularly relevant. In the first, vulnerabilities remain unknown when the artifact is produced. Although the project may appear functionally plausible, its security status is unclear. Securing such artifacts requires vulnerability detection, repair guided by detection outputs, and post-repair verification. The usefulness of this process depends in part on whether detection reports are reliable enough to support downstream repair.

In the second situation, vulnerabilities have already been identified through exploit confirmation, prior analysis, or benchmark-defined targets. In this case, the focus shifts from discovery to mitigation. The key issue is whether repair can remove the reported vulnerability without causing regression, incomplete fixes, or other side effects revealed during verification. This makes repair trustworthiness a central concern.

Figure \ref{fig:motivating_scenarios} shows these two scenarios. Although they begin differently, both follow the same detect–repair–verify backbone and are evaluated against the same endpoint: whether the resulting artifact is both functionally correct and security-preserving. This common structure motivates the empirical study presented next.

\section{Methodology}
\label{methodology}

\begin{figure}
    \centering
    \includegraphics[width=1\linewidth]{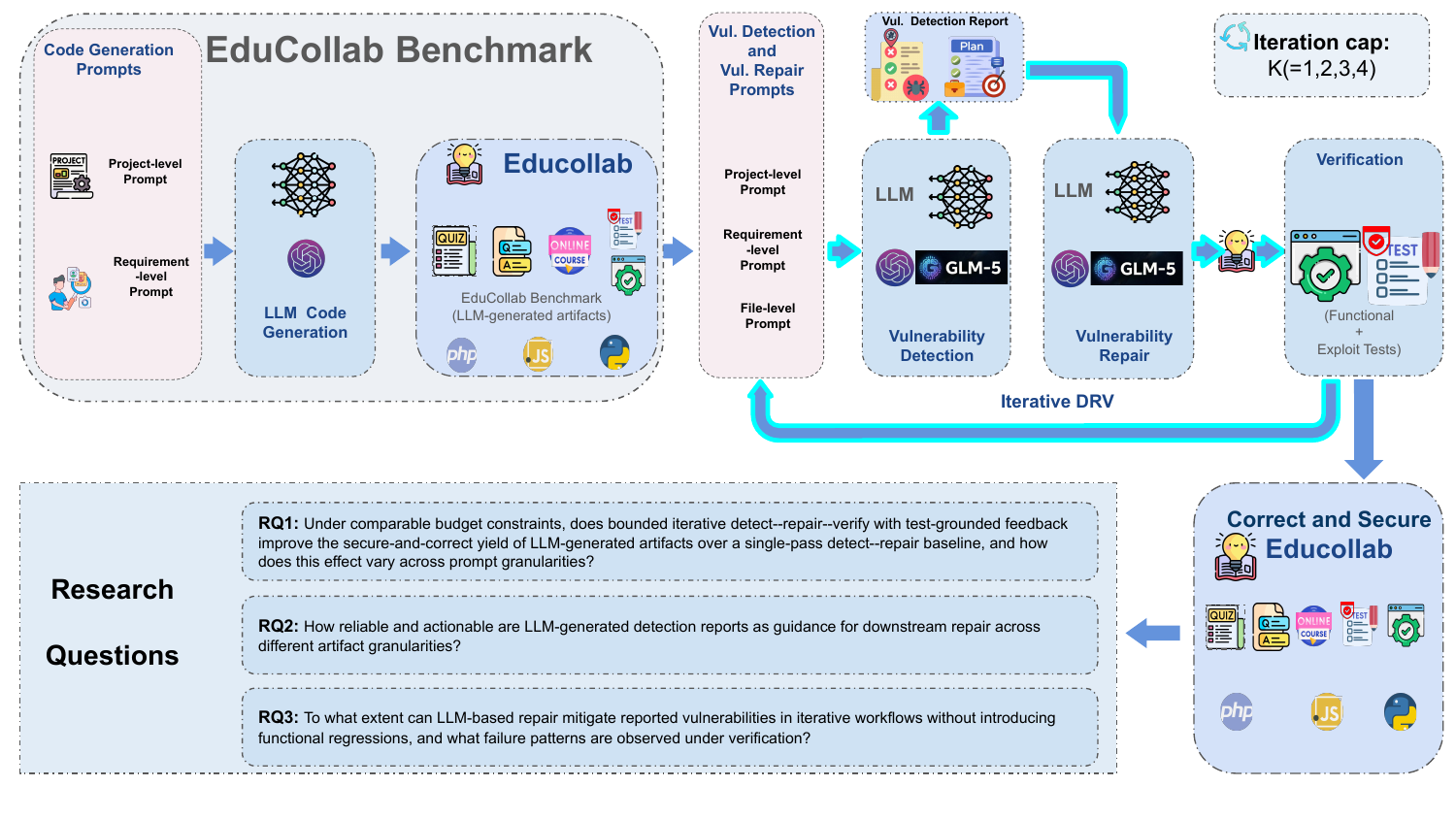}
    \caption{Overview of the experimental pipeline. Starting from the EduCollab benchmark (PHP/JS/Python), the study evaluates artifacts under three prompt granularities: project-level, requirement-level, and file-level. These artifacts are processed through a detect--repair--verify workflow, where repair generates LLM-based patches and verification reruns the executable test suites. The resulting feedback supports bounded iteration and enables the evaluation of RQ1--RQ3.}
    \label{fig:methodology}
\end{figure}

This section describes the study methodology as the end-to-end pipeline shown in Fig.~\ref{fig:methodology}. The methodology has two parts. First, a multi-language EduCollab benchmark is constructed from LLM-generated artifacts under different code-generation prompt granularities and paired with functional and exploit test suites. Second, this benchmark is used to evaluate vulnerability detection and repair under different prompt settings and workflow conditions, with each repaired artifact verified by rerunning the corresponding exploit target together with the functional test suite. Outcomes are recorded per target and then aggregated by model, language, granularity, and workflow setting for the analyses in RQ1--RQ3.

\subsection{Benchmark Construction}

The benchmark is built around a shared EduCollab functional blueprint instantiated in three language ecosystems: PHP, JavaScript, and Python. All implementations follow an MVC-style architecture and are course-centric, with \textit{Course} serving as the tenant boundary. Rather than using manually developed reference systems, this study defines the benchmark as a set of runnable LLM-generated artifacts produced under controlled prompting conditions. Project-level prompts are used to generate complete project artifacts, whereas requirement-level prompts are used to generate artifacts incrementally through requirement-scoped development steps. These artifacts serve as the primary subjects of the detect--repair--verify experiments.

The benchmark is constructed in four stages. First, a common functional specification is defined together with risk-guided use-case selection. Second, project-level and requirement-level prompt settings are designed for artifact generation. Third, runnable artifacts are generated and selected across languages. Fourth, each artifact is paired with functional and exploit test suites for evaluation. This design keeps application semantics aligned across languages and supports controlled comparison under shared workflows, use cases, and verification conditions. The benchmark and accompanying artifacts are available in a public repository.\footnote{\url{https://github.com/Hahappyppy2024/EmpricalVDR}} In the subsequent experiments, LLMs are involved in two ways: artifact generation under the two prompt settings, and patch generation during repair, where model-produced edits transform baseline artifacts into post-repair artifacts evaluated with executable tests.

\subsubsection{Stage~1: Functional Specification and Risk-Guided Use-Case Selection}
The benchmark is built around a shared functional specification to support controlled comparison across languages. Three EduCollab implementations in PHP, JavaScript, and Python are developed against the same blueprint, so that differences observed in detection, repair, and verification can be attributed to workflow behavior and language or runtime effects rather than differences in application scope. The specification captures representative workflows of a course-centric collaboration platform, including authentication and session management, course and membership management with role-based access control, token-based invite joining, content posting and search, file submission and resource distribution, and administrative functions such as auditing and data import and export.

The selection of use cases is guided not only by application realism, but also by security relevance. In particular, the implemented use cases are chosen to expose representative attack surfaces aligned with the OWASP Top~10:2025 risk taxonomy (\texttt{R01--R10})~\citep{owasp_top10_2025}. This design does not aim to exhaustively cover all web security issues. Instead, it provides a controlled set of functionally meaningful application behaviors whose security exposure can be related to widely used risk families, allowing evaluation and reporting to remain comparable across languages.

Table~\ref{tab:functional_features_usecases} lists the concrete use cases instantiated in the projects. Several of these use cases are security-relevant by design, including rendering and search surfaces, ZIP and CSV import and download, path handling, misconfiguration checks, audit logging, and exceptional-condition handling. These features provide realistic attack surfaces commonly encountered in web application security evaluation. All three implementations expose the same set of use cases and are validated with a shared functional test suite to establish semantic equivalence before security testing and the subsequent LLM-driven detect, repair, and verify experiments.

Table~\ref{tab:top10_summary_ranked} summarizes the OWASP Top~10:2025 categories used in this study, including the category name, identifier, and operational interpretation adopted for test design and result reporting. Table~\ref{tab:feature_rank_2025_sorted} then links the benchmark's functional use cases to these risk categories. Multiple use cases may map to the same category because a single risk family can arise in different parts of the application, such as authorization checks in both membership management and role updates. The mapping covers nine categories (\texttt{R01}, \texttt{R02}, \texttt{R04--R10}). \texttt{R03}, which concerns software supply chain failures, is driven primarily by dependency choices and build or deployment processes, and therefore falls outside the in-application use cases covered by this benchmark.

\begin{table}
\centering
\caption{Functional use cases instantiated in EduCollab (shared across PHP/JS/Python).}
\label{tab:functional_features_usecases}
\begin{tblr}{
  row{1} = {font=\bfseries},
  hlines = {dotted},
  hline{1,16} = {-}{solid,0.08em},
  hline{2} = {-}{solid,0.05em},
}
ID  & Use case                                                                           \\
U1  & Course and membership management (RBAC).                                           \\
U2  & Role update (member role change).                                                  \\
U3  & Security misconfiguration checks (cookie flags, \texttt{nosniff}, debug leakage).  \\
U4  & Invite-link join (token-based enrollment).                                         \\
U5  & Post/comment and search (keyword, sort).                                           \\
U6  & Stored/reflected rendering surfaces (XSS-relevant outputs).                        \\
U7  & ZIP/CSV import and download (CSV formula surface).                                 \\
U8  & Self-escalation via role change (assistant promotes self to admin).                \\
U9  & Authentication and session lifecycle (login/logout).                               \\
U10 & Assignment submission with upload (file handling).                                 \\
U11 & Upload/download with path-handling constraints (integrity-relevant surfaces).      \\
U12 & ZIP/CSV import and download (integrity checks).                                    \\
U13 & Admin audit-log view (\texttt{/admin/audit}); missing logging for critical events. \\
U14 & Exceptional-condition handling (malformed inputs; error leakage surfaces).         
\end{tblr}
\end{table}

\begin{table}
\centering
\caption{2025 Top-10 web application security risks.}
\label{tab:top10_summary_ranked}
\begin{tblr}{
  row{1} = {c,font=\bfseries},
  cell{2}{1} = {c},
  cell{3}{1} = {c},
  cell{4}{1} = {c},
  cell{5}{1} = {c},
  cell{6}{1} = {c},
  cell{7}{1} = {c},
  cell{8}{1} = {c},
  cell{9}{1} = {c},
  cell{10}{1} = {c},
  cell{11}{1} = {c},
  hlines = {dotted},
  hline{1-2,12} = {-}{solid},
}
Rank & Category (ID)                                             & What it means                                                                                                 \\
1    & Broken Access Control (R01)                               & {Authorization checks are missing or bypassable,\\so users can access data/actions beyond their permissions.} \\
2    & Security Misconfiguration (R02)                           & {The app, server, framework, or cloud settings~\\are insecure or left in unsafe defaults.}                    \\
3    & Software Supply Chain Failures (R03)                      & {Risks come from dependencies, build artifacts,\\~or delivery pipelines rather than your own code.}           \\
4    & Cryptographic Failures (R04)                              & {Data is not properly protected because cryptography\\~is missing, weak, or misused.}                         \\
5    & Injection (R05)                                           & {Untrusted input is interpreted as code/commands/queries\\~by an interpreter or engine.}                      \\
6    & Insecure Design (R06)                                     & {The system’s design lacks necessary security controls~\\or abuse-resistance from the start.}                 \\
7    & Authentication Failures (R07)                             & {Identity and session mechanisms are weak, allowing~\\account takeover or session abuse.}                     \\
8    & Software or Data Integrity Failures (R08)                 & {The app trusts code/data that can be tampered with,\\leading to unsafe updates or corrupted flows.}          \\
9    & Security Logging \textbackslash{} Alerting Failures (R09) & {Security-relevant events are not monitored well,~\\delaying detection and response.}                         \\
10   & Mishandling of Exceptional Conditions (R10)               & {Error/exception paths are unsafe and can leak\\~information, bypass checks, or fail open.}                   
\end{tblr}
\end{table}

\begin{table}[t]
\centering
\caption{Functional features (use cases) aligned with the ranked Top-10 risk list (R01--R10), sorted by ascending rank. The table covers 9 of the 10 risk categories (R01, R02, R04--R10); R03 (software supply chain failures) is treated as a process-level risk and thus not represented as an in-app use case.}
\label{tab:feature_rank_2025_sorted}
\begin{tblr}{
  row{1} = {c,font=\bfseries},
  cell{2}{1} = {c},
  cell{3}{1} = {c},
  cell{4}{1} = {c},
  cell{5}{1} = {c},
  cell{6}{1} = {c},
  cell{7}{1} = {c},
  cell{8}{1} = {c},
  cell{9}{1} = {c},
  cell{10}{1} = {c},
  cell{11}{1} = {c},
  cell{12}{1} = {c},
  cell{13}{1} = {c},
  cell{14}{1} = {c},
  cell{15}{1} = {c},
  hlines = {dotted},
  hline{1-2,16} = {-}{solid},
}
\textbf{Rank ID} & \textbf{Functional feature (use case)} \\
R01 & Course \textbackslash{} Membership (RBAC) \\
R01 & Role Update (member role change) \\

R02 & Security Misconfiguration (cookie flags / nosniff / debug leakage) \\

R04 & Invite Link Join (token-based join) \\

R05 & Post/Comment \textbackslash{} Search (keyword/sort) \\
R05 & Stored/Reflected Rendering (XSS surface) \\
R05 & ZIP/CSV Import \textbackslash{} Download (CSV formula) \\

R06 & Self-escalation via Role Change (assistant can promote self to admin) \\

R07 & Auth \textbackslash{} Session (login/logout) \\

R08 & Assignment Submission Upload (file handling) \\
R08 & Upload/Download \textbackslash{} Path Handling (integrity surfaces) \\
R08 & ZIP/CSV Import \textbackslash{} Download (integrity) \\

R09 & Admin Audit Log View (/admin/audit) with missing logging for critical events \\

R10 & Exceptional Conditions (malformed inputs / error leakage) \\
\end{tblr}
\end{table}

\subsubsection{Stage~2: Prompt Granularity Design for Artifact Generation}

Based on the shared functional blueprint and use cases described above, this stage defines two prompt granularities for artifact generation: project-level and requirement-level. Both are designed to produce runnable LLM-generated artifacts under the same application semantics, but they differ in how much system context is provided to the model at each generation step. Project-level prompts present the system as a whole, whereas requirement-level prompts provide the specification incrementally through requirement-scoped steps. This distinction allows the benchmark to capture artifacts generated under different prompt scopes while preserving a common functional target across settings.

\paragraph{Project-level.}
Project-level prompts provide a global specification of the system. They include the tenant boundary (\texttt{course}), user roles, major modules, core workflows, data and access-control expectations, and the overall technology stack. These elements are included so that the model receives enough system-wide context to generate a complete implementation with consistent structure and interactions across components. The goal of this setting is to produce a complete runnable artifact from a single prompt.

\paragraph{Requirement-level.}
Requirement-level prompts are organized around the five requirement blocks in Table~\ref{tab:requirement_decomposition}. Each prompt focuses on one requirement block and includes the functional scope of that requirement, the interfaces and behaviors that must remain consistent with previously generated code, and the constraints needed to preserve earlier functionality. These elements are included so that new functionality can be added incrementally without breaking the existing implementation. The goal of this setting is to produce a runnable artifact after each requirement step, rather than only at the end of the full sequence.

\begin{table}
\centering
\caption{Requirement-level functional decomposition used for incremental artifact generation.}
\label{tab:requirement_decomposition}
\begin{tblr}{
  row{1} = {font=\bfseries},
  column{1} = {c},
  hlines = {dotted},
  hline{1-2,7} = {-}{solid},
}
Req. & Functional Scope                                                & Main Modules                                & Representative APIs / Features                                                                                                                                    \\
R1   & {Bootstrapping, authentication,\\~user, and course management}  & User, Course                                              & {DB initialization and admin seeding;~\\registration/login/logout/me; user listing;~\\course CRUD}                                                                \\
R2   & {Membership management~\\and course-level role enforcement}     & Membership                                                & {Membership add/list/update/remove;\\~user membership listing; access-control\\~helpers such as requireLogin,~\\requireCourseMember,\\~and requireTeacherOrAdmin} \\
R3   & {Post, comment,\\~and search workflows}                         & Post, Comment                                             & {Post CRUD; comment CRUD;\\~keyword search over posts/comments;~\\role-scoped discussion workflows}                                                               \\
R4   & {Assignment, submission,~\\and upload management}               & {Assignment, Submission,~\\Upload}                        & {Assignment CRUD;\\~student submission create/update/list;~\\staff-side submission listing;\\~file upload/list/download/delete}                                   \\
R5   & {Question bank,~\\quiz management,~\\and student quiz attempts} & {Question, Quiz, QuizQuestion,\\~QuizAttempt, QuizAnswer} & {Question bank CRUD; quiz CRUD;~\\quiz-question configuration;\\~attempt start, answer submission,~\\attempt submission, and self-view}                           
\end{tblr}
\end{table}

\subsubsection{Stage~3: Cross-language Artifact Generation and Selection}

Based on the shared functional specification in Stage~1 and the prompt granularities defined in Stage~2, EduCollab artifacts are generated in three language ecosystems: PHP, JavaScript, and Python. Generation is carried out with multiple LLMs, including ChatGPT~\citep{openai_chatgpt_2026}, DeepSeek~\citep{deepseek_platform_2026}, and GLM~\citep{zhipu_glm_docs_2026}, under the same prompt design so that artifact construction can be examined under comparable functional scope across languages.

For each language and prompt granularity, the goal is to produce a runnable artifact that follows the shared EduCollab blueprint, including course-centric tenancy, role-based access control, and the required application workflows. The generated artifacts are expected to expose comparable resource types, such as courses, memberships, posts, uploads, assignments, and submissions, and, when applicable, to support both Web UI workflows and corresponding \texttt{/api/*} endpoints. This alignment is needed for applying common functional and exploit tests across implementations.

In practice, artifact quality varies substantially across models. Although ChatGPT~\citep{openai_chatgpt_2026}, DeepSeek~\citep{deepseek_platform_2026}, and GLM~\citep{zhipu_glm_docs_2026} are all used in generation attempts, only ChatGPT consistently produces artifacts that satisfy the executability and completeness requirements for inclusion in the final benchmark. Artifacts generated by DeepSeek and GLM do not meet the same requirements and are therefore not retained. The final benchmark used in this study thus consists of runnable ChatGPT-generated artifacts under the project-level and requirement-level prompt settings across the three language ecosystems.

\subsubsection{Stage~4: Functional and Exploit Test Suites}

This stage turns the benchmark into an executable benchmark by pairing each artifact with two complementary test suites. The \emph{functional} test suite checks functional correctness, while the \emph{exploit} test suite checks whether targeted vulnerability conditions can be triggered at high-risk surfaces. Together, they provide test-grounded evidence for both correctness and security during evaluation.

\paragraph{Functional correctness tests.}
Functional tests are organized according to artifact granularity. For project-level and file-level artifacts, the tests are derived from the shared functional use cases listed in Table~\ref{tab:functional_features_usecases}. For requirement-level artifacts, the tests are aligned with the requirement blocks in Table~\ref{tab:requirement_decomposition}, so that each intermediate artifact is checked against the functionality expected at its corresponding requirement step. Across settings, these tests exercise the intended workflows of the course-centric collaboration system, including authentication and session handling, course creation and membership management, invite-based joining, posting and search, assignment submission with uploads, and data import and export. They are executed under appropriate authorization contexts so that intended actions succeed for authorized roles and are rejected for unauthorized roles, in accordance with the shared functional specification.

\paragraph{Exploit tests.}
Exploit tests are constructed separately for project-level, requirement-level, and file-level artifacts. For each granularity, the test targets are defined by the vulnerabilities listed in Table~\ref{tab:project_level_vulns}, Table~\ref{tab:requirement_level_vulns}, and Table~\ref{tab:file_level_vulns}, respectively. Each exploit test is tied to a specific vulnerability entry and is designed to check whether the corresponding insecure behavior can be triggered in the artifact under test. The suite therefore does not serve as a generic security scan; instead, it operationalizes vulnerability verification as a set of targeted, adversarial checks with clear pass-or-fail outcomes. Across the three granularities, these tests cover multiple OWASP Top~10-related vulnerability types, including broken access control, authentication weaknesses, injection-related flaws, security misconfiguration, and integrity failures. By using crafted inputs and boundary-violating requests, the exploit tests determine whether the observed behavior demonstrates the targeted vulnerability.

\begin{table}
\centering
\caption{Project-level vulnerabilities included in the benchmark.}
\label{tab:project_level_vulns}
\begin{tblr}{
  column{4} = {c},
  row{1} = {font=\bfseries},
  hline{1-2} = {-}{0.08em},
  hline{8,17} = {-}{dotted},
  hline{25} = {-}{0.08em},
}
Language   & Vul. ID & Vulnerability Name                                & Top-10 Type \\
JavaScript & VUL001       & BAC Static Exposure uploads public download       & 1 \\
JavaScript & VUL002       & IDOR Student Deletes Another Student Comment      & 1 \\
JavaScript & VUL003       & IDOR Student Edits Another Student Post           & 1 \\
JavaScript & VUL004       & IDOR Submission Overwrite                         & 1 \\
JavaScript & VUL005       & Info Disclosure Any Logged In User Can List Users & 1 \\
JavaScript & VUL006       & Upload HTML same origin active content            & 8 \\
Python     & VUL001  & default admin credentials                         & 7 \\
Python     & VUL002  & user list access control                          & 1 \\
Python     & VUL003  & course edit delete BAC                            & 1 \\
Python     & VUL004  & cross course membership IDOR                      & 1 \\
Python     & VUL005  & post IDOR                                         & 1 \\
Python     & VUL006  & comment IDOR                                      & 1 \\
Python     & VUL007  & question answer leak                              & 1 \\
Python     & VUL008  & upload path traversal                             & 8 \\
Python     & VUL009  & quiz answer tampering                             & 8 \\
PHP        & VUL001  & default admin credentials                         & 7 \\
PHP        & VUL002  & session fixation                                  & 7 \\
PHP        & VUL003  & CSRF                                              & 1 \\
PHP        & VUL004  & cross course membership IDOR                      & 1 \\
PHP        & VUL005  & post IDOR                                         & 1 \\
PHP        & VUL006  & comment IDOR                                      & 1 \\
PHP        & VUL007  & question bank answer leak                         & 1 \\
PHP        & VUL008  & quiz answer tampering                             & 8 \\
\end{tblr}
\end{table}

\begin{table}
\centering
\caption{Requirement-level vulnerabilities included in the benchmark.}
\label{tab:requirement_level_vulns}
\begin{tblr}{
  column{4} = {c},
  row{1} = {font=\bfseries},
  hline{1-2,44} = {-}{0.08em},
  hline{15,33} = {-}{dotted},
}
Language   & Req. & Vul. ID & Vulnerability Name                  & Top-10 Type \\
JavaScript & R1   & VUL001  & hardcoded admin                     & 7 \\
JavaScript & R1   & VUL002  & session fixation                    & 7 \\
JavaScript & R1   & VUL003  & course IDOR                         & 1 \\
JavaScript & R1   & VUL004  & CSRF                                & 1 \\
JavaScript & R1   & VUL005  & default session secret              & 2 \\
JavaScript & R2   & VUL001  & roster disclosure                   & 1 \\
JavaScript & R2   & VUL002  & membership CSRF                     & 1 \\
JavaScript & R3   & VUL001  & post IDOR                           & 1 \\
JavaScript & R3   & VUL002  & comment IDOR                        & 1 \\
JavaScript & R4   & VUL001  & upload path traversal               & 8 \\
JavaScript & R4   & VUL002  & submission attachment IDOR          & 1 \\
JavaScript & R5   & VUL001  & question answer exposure            & 1 \\
JavaScript & R5   & VUL002  & quiz question injection             & 5 \\
Python     & R1   & VUL001  & default admin                       & 7 \\
Python     & R1   & VUL002  & forged session                      & 7 \\
Python     & R1   & VUL003  & user list BAC                       & 1 \\
Python     & R1   & VUL004  & course IDOR                         & 1 \\
Python     & R1   & VUL005  & CSRF                                & 1 \\
Python     & R2   & VUL005  & membership roster BAC               & 1 \\
Python     & R2   & VUL006  & membership CSRF                     & 1 \\
Python     & R3   & VUL001  & post IDOR                           & 1 \\
Python     & R3   & VUL002  & comment IDOR                        & 1 \\
Python     & R3   & VUL003  & discussion CSRF                     & 1 \\
Python     & R4   & VUL001  & upload access BAC                   & 1 \\
Python     & R4   & VUL002  & submission attachment logic         & 8 \\
Python     & R4   & VUL003  & upload storage path leak            & 2 \\
Python     & R4   & VUL004  & assignment CSRF                     & 1 \\
Python     & R5   & VUL001  & quiz window bypass                  & 1 \\
Python     & R5   & VUL002  & quiz attempt tampering              & 8 \\
Python     & R5   & VUL003  & quiz CSRF                           & 1 \\
PHP        & R1   & VUL001  & hardcoded admin credentials         & 7 \\
PHP        & R1   & VUL002  & session fixation                    & 7 \\
PHP        & R1   & VUL003  & user directory exposure             & 1 \\
PHP        & R1   & VUL004  & course IDOR                         & 1 \\
PHP        & R1   & VUL005  & CSRF missing                        & 1 \\
PHP        & R2   & VUL003  & CSRF html forms                     & 1 \\
PHP        & R3   & VUL001  & post IDOR                           & 1 \\
PHP        & R3   & VUL002  & comment IDOR                        & 1 \\
PHP        & R4   & VUL001  & insecure file upload                & 8 \\
PHP        & R4   & VUL002  & submission attachment IDOR          & 1 \\
PHP        & R5   & VUL001  & quiz answer exposure                & 1 \\
PHP        & R5   & VUL002  & quiz window bypass                  & 1 \\
\end{tblr}
\end{table}

\begin{table}
\centering
\caption{File-level vulnerabilities included in the benchmark.}
\label{tab:file_level_vulns}
\begin{tblr}{
  column{4} = {c},
  row{1} = {font=\bfseries},
  hline{1-2,22} = {-}{0.08em},
  hline{8,16} = {-}{dotted},
}
Language   & Vul. ID & Vulnerability Name                                & Top-10 Type \\
JavaScript & VUL001  & {POC BAC Static Exposure uploads~\\public download}       & 1 \\
JavaScript & VUL002  & {POC IDOR Student Deletes Another\\~Student Comment}      & 1 \\
JavaScript & VUL003  & {POC IDOR Student Edits Another\\~Student Post}           & 1 \\
JavaScript & VUL004  & POC IDOR Submission Overwrite                             & 1 \\
JavaScript & VUL005  & {POC Info Disclosure Any Logged\\~In User Can List Users} & 1 \\
JavaScript & VUL006  & {POC Upload HTML same origin\\~active content}            & 8 \\
Python     & VUL001  & session fixation                                          & 7 \\
Python     & VUL002  & cross course membership IDOR                              & 1 \\
Python     & VUL003  & unauthorized post edit                                    & 1 \\
Python     & VUL004  & unauthorized comment edit                                 & 1 \\
Python     & VUL005  & question answer key exposure                              & 1 \\
Python     & VUL006  & cross course quiz attempt IDOR                            & 1 \\
Python     & VUL007  & foreign question attachment                               & 8 \\
PHP        & VUL001  & session fixation                                          & 7 \\
PHP        & VUL002  & cross course membership IDOR                              & 1 \\
PHP        & VUL003  & unauthorized post edit                                    & 1 \\
PHP        & VUL004  & unauthorized comment edit                                 & 1 \\
PHP        & VUL005  & question answer key exposure                              & 1 \\
PHP        & VUL006  & cross course quiz attempt IDOR                            & 1 \\
PHP        & VUL007  & foreign question attachment                               & 8 \\
\end{tblr}
\end{table}

To improve readability, Table~\ref{tab:abbreviation_summary} explains the abbreviations used in vulnerability names across the benchmark.

\begin{table}[t]
\centering
\caption{Abbreviations used in vulnerability names and workflow descriptions.}
\label{tab:abbreviation_summary}
\begin{tblr}{
  row{1} = {font=\bfseries},
  colspec = {Q[l,wd=2.0cm] Q[l,wd=4.2cm] Q[l,wd=8.0cm]},
  hline{1-2,10} = {-}{0.08em},
  hline{3-9} = {-}{dotted},
}
Abbreviation & Full form & Meaning in this study \\
POC  & Proof of Concept & Indicates a concrete exploit-oriented vulnerability instance used as a file-level target. \\
IDOR & Insecure Direct Object Reference & A broken access control flaw in which an attacker can access or modify another user's object by manipulating identifiers. \\
BAC  & Broken Access Control & Access restrictions are missing, bypassable, or incorrectly enforced. \\
CSRF & Cross-Site Request Forgery & A vulnerability in which unauthorized state-changing requests are triggered through an authenticated user's browser. \\
CRUD & Create, Read, Update, Delete & The standard set of resource-management operations used to describe application functionality. \\
DRV  & Detect--Repair--Verify & The workflow used in the study to detect vulnerabilities, generate repairs, and verify patched artifacts. \\
SCY  & Secure-and-Correct Yield & The proportion of artifacts or targets that pass both functional and exploit-based verification. \\
LLM  & Large Language Model & A generative model used for artifact generation, vulnerability detection, and repair. \\
\end{tblr}
\end{table}

In total, the benchmark contains 3 project-level artifacts, 15 requirement-level artifacts, and 20 file-level artifacts. Under the artifact-level counting scheme used in the evaluation, each artifact is paired with one corresponding functional test suite for correctness validation. This yields 15, 15, and 15 functional test suites for the project-, requirement-, and file-level settings, respectively. The corresponding exploit-test counts are 23, 42, and 20, following the vulnerability sets summarized in Table~\ref{tab:project_level_vulns}, Table~\ref{tab:requirement_level_vulns}, and Table~\ref{tab:file_level_vulns}. Table~\ref{tab:dataset_scale_by_granularity} summarizes the resulting benchmark scale by granularity.

\begin{table}[t]
\centering
\caption{Benchmark statistics across project-level, requirement-level, and file-level settings.}
\label{tab:dataset_scale_by_granularity}
\begin{tblr}{
  row{1} = {font=\bfseries},
  column{2-4} = {c},
  hline{1-2,5} = {-}{0.08em},
  hline{3-4} = {-}{dotted},
}
Granularity & \#Artifacts & \#Functional Test Suites & \#Exploit Tests \\
Project-level     & 3  & 15  & 23 \\
Requirement-level & 15 & 15 & 42 \\
File-level        & 20 & 15 & 20 \\
\end{tblr}
\end{table}

\subsection{Detect--Repair--Verify Workflows and Verification-Based Measures}

This section describes how detect--repair--verify evaluation is carried out on the multi-language EduCollab benchmark. Two settings are considered. In the first, vulnerabilities are latent in the artifact and must be identified before repair can proceed. In the second, vulnerabilities are already specified through exploit targets, so the workflow begins from a pre-identified repair target. Although the two settings differ in how they enter the workflow, both follow the same bounded iterative structure of detection, repair, and verification. The remainder of this section introduces these two workflow settings, explains the model-selection strategy used across prompt granularities, and then defines the verification-based measures used in RQ1--RQ3.

\subsubsection{Workflow for Artifacts with Latent Vulnerabilities}
\begin{figure}[H]
    \centering
    \includegraphics[width=\linewidth]{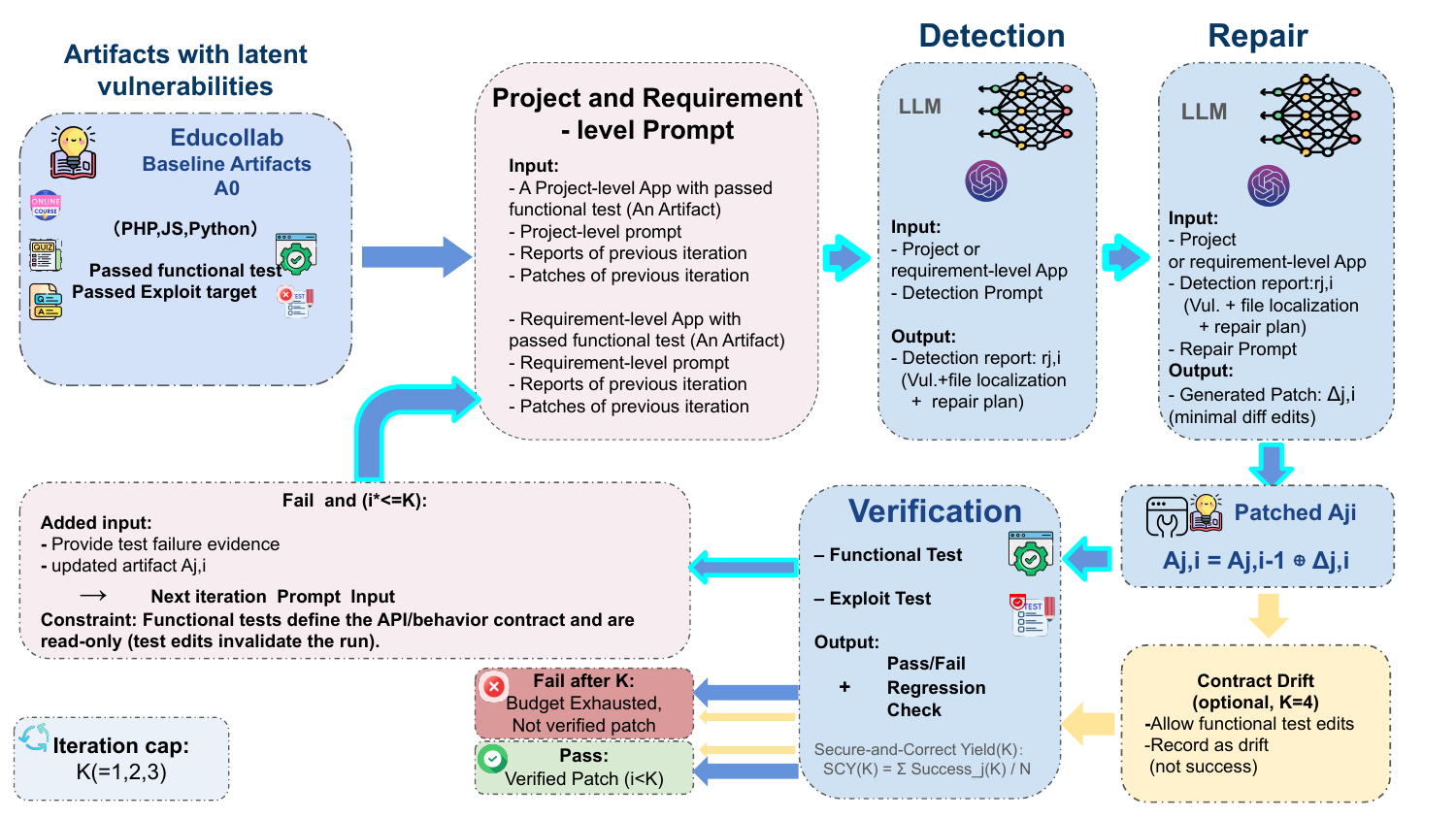}
    \caption{Detect--repair--verify workflow for artifacts with latent vulnerabilities. Starting from a runnable baseline artifact, the workflow uses project-level or requirement-level prompts to detect candidate vulnerabilities, generate patches, and verify patched artifacts by rerunning both functional and exploit tests under a bounded iteration budget.}
    \label{fig:latent_workflow}
\end{figure}

Figure~\ref{fig:latent_workflow} shows the workflow used for artifacts with latent vulnerabilities. This setting starts from a baseline artifact $A_0$ that passes the functional test suite and is treated as a runnable project artifact whose vulnerability status is not given to the model in advance. The workflow is applied to project-level and requirement-level artifacts, where the goal is to detect candidate vulnerabilities, generate repairs, and verify whether the patched artifact is both functionally correct and no longer vulnerable under the exploit test.

At each iteration, the model receives the current project- or requirement-level artifact together with the corresponding prompt, and, when available, the reports and patches from previous iterations. In the detection stage, the model produces a structured report $r_{j,i}$ that identifies the vulnerability, localizes the relevant code, and proposes a repair plan. The repair stage then uses the artifact, the detection report, and the repair prompt to generate a patch $\Delta_{j,i}$. This patch is applied to the current artifact to obtain the updated version $A_{j,i}$.

The patched artifact is then evaluated in the verification stage. Functional tests are rerun to check whether intended behavior is preserved, and exploit tests are rerun to determine whether the target vulnerability is still reproducible. Verification therefore yields both a security result and a regression check. If both checks succeed, the case is recorded as a verified patch. Otherwise, the workflow proceeds to the next iteration, up to the iteration cap $K$. When no verified patch is obtained within the budget, the case is recorded as budget exhausted. An optional contract-drift branch is also included at $K=4$: functional test edits may be allowed to investigate drift, but such cases are recorded as drift rather than success.

\subsubsection{Workflow for Artifacts with Pre-identified Vulnerabilities}
\begin{figure}[H]
    \centering
    \includegraphics[width=\linewidth]{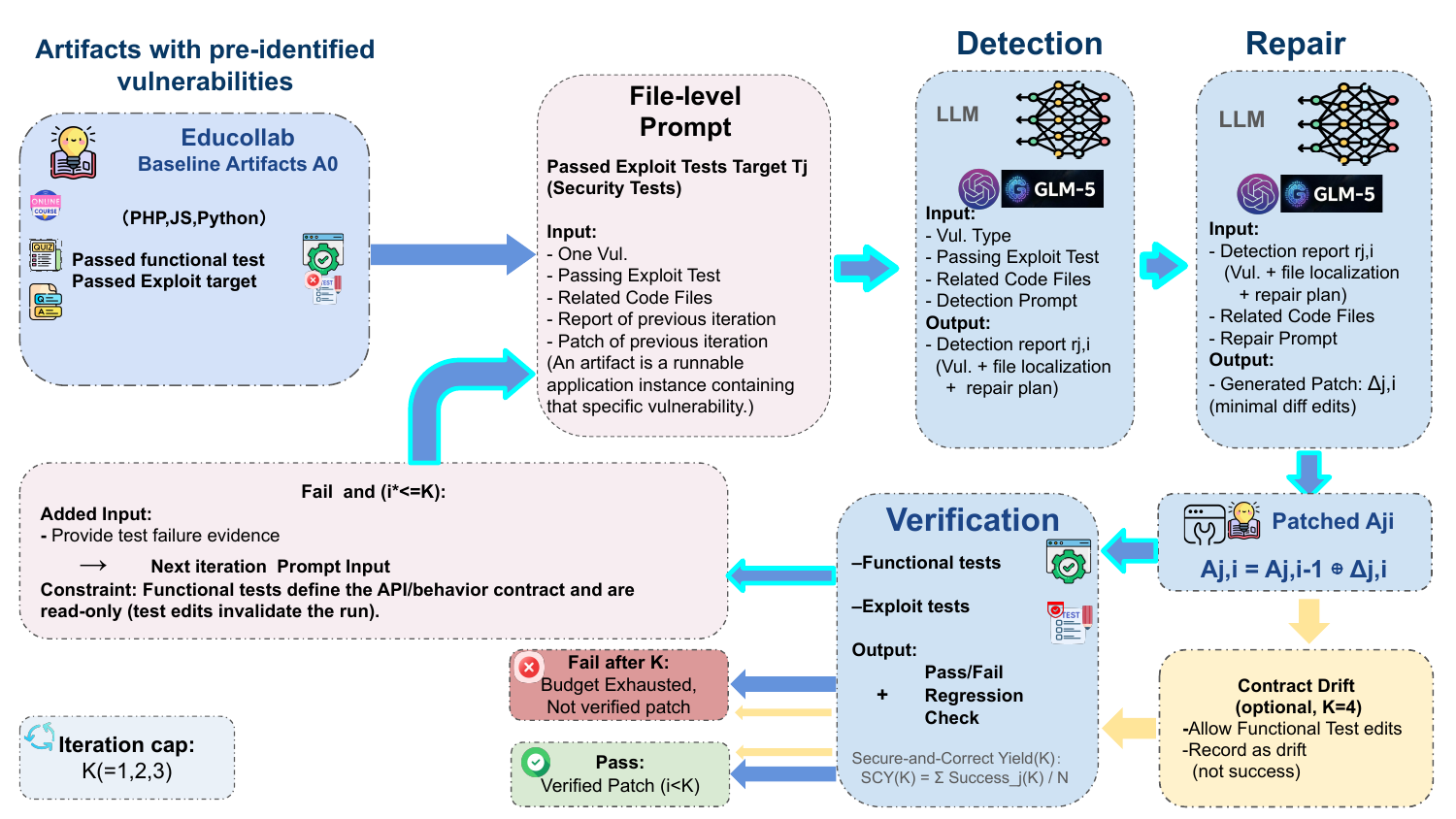}
    \caption{Detect--repair--verify workflow for artifacts with pre-identified vulnerabilities. Starting from a runnable baseline artifact paired with a specific exploit target, the workflow uses file-level prompts to localize and repair one vulnerability at a time, and verifies each patched artifact by rerunning both functional and exploit tests under a bounded iteration budget.}
    \label{fig:preidentified_workflow}
\end{figure}

Figure~\ref{fig:preidentified_workflow} illustrates the workflow used for artifacts with pre-identified vulnerabilities. This setting starts from a baseline artifact $A_0$ that passes the functional test suite and is paired with a specific exploit target $T_j$. Each artifact is therefore a runnable application instance for which the target vulnerability has already been identified and can be checked through its corresponding exploit test.

The workflow is instantiated with file-level prompts. At each iteration, the model receives the target vulnerability, the passing exploit test, the related code files, and, when applicable, the report and patch from the previous iteration. Based on this input, the detection stage produces a structured report that localizes the vulnerability and proposes a repair plan. The repair stage then generates a minimal patch $\Delta_{j,i}$, which is applied to the artifact to obtain the updated version $A_{j,i}$.

The patched artifact is evaluated in the verification stage using both functional tests and exploit tests. Functional tests are rerun to check whether the patch preserves intended behavior, while the exploit test is rerun to determine whether the target vulnerability has been mitigated. Verification therefore yields both a security result and a regression check. If the artifact passes both checks, the patch is counted as a verified success. Otherwise, the workflow continues to the next iteration, up to the iteration cap $K$. When no verified patch is obtained within the budget, the case is recorded as budget exhausted. An optional contract-drift branch is also included at $K=4$: functional test edits may be allowed to investigate drift, but such cases are recorded as drift rather than success.

\subsubsection{Large Language Model Selection}
Model selection is designed to balance controlled evaluation with feasible cross-model comparison. For project-level and requirement-level settings, ChatGPT-5 is used as the sole model. These settings involve larger application context and broader reasoning over system structure, making a single model choice preferable for stable comparison. For file-level settings, both ChatGPT-5 and GLM-5 are evaluated. Because file-level prompts operate on a narrower context, they provide a more controlled setting for cross-model comparison under the same benchmark and test-grounded detect--repair--verify protocol.

\subsubsection{Verification-Based Measures}
All workflows are evaluated under the same verification logic. Functional correctness is assessed with the corresponding functional test suite, exploitability is assessed with the corresponding security exploit tests, and a case is treated as secure-and-correct only when it satisfies both conditions. These verification outcomes provide the common basis for the RQ-specific measures reported in Section~\ref{Results}.

\section{Results}
\label{Results}

\subsection{RQ1 (Pipeline-level effectiveness).}

\subsubsection{Measures}
RQ1 examines the effectiveness of bounded detect--repair--verify workflows under comparable repair budgets. The comparison covers three repair scopes: project-level, requirement-level, and file-level, and contrasts single-pass repair with iterative repair.

Four indicators are used throughout. \emph{Used Iterations} records the iteration at which the reported result is obtained. \emph{Func.\ Pass Rate} shows whether the repaired artifact still preserves the expected functional behavior. \emph{Exploit Success Rate} shows whether the corresponding exploit remains successful after repair, where lower values indicate better security improvement. \emph{S\textbackslash{}C Yield Rate} records whether a case is both functionally correct and non-exploitable, and serves as the main indicator of secure-and-correct convergence.

Table~\ref{tab:rq1_pipeline_effectiveness} reports the project-level and requirement-level results. At the project level, results are summarized per language for the full R1--R5 artifact. At the requirement level, results are reported separately for each requirement to show how workflow effectiveness changes when the repair target is narrowed. Tables~\ref{tab:file_level_result_chatgpt} and~\ref{tab:file_level_result_glm} report the same indicators at the file level for ChatGPT-5 and GLM-5. Together, these tables show how repair effectiveness varies with repair scope, workflow setting, and model.

\subsubsection{Findings.}

\paragraph{Project-level observations.}
Table~\ref{tab:rq1_pipeline_effectiveness} shows that project-level repair is the least stable setting under the bounded D--R--V workflow. JavaScript reaches a secure-and-correct outcome in a single pass, so iteration is not needed. PHP and Python both fail in the single-pass setting. Python improves with iteration and reaches a secure-and-correct outcome at iteration 3, while PHP does not converge within the bounded budget. This suggests that iterative D--R--V can help at the project level, but the effect is not consistent across languages, and convergence remains difficult when the repair scope is broad.

\paragraph{Requirement-level observations.}
Table~\ref{tab:rq1_pipeline_effectiveness} also shows that requirement-level repair is more stable than project-level repair, although the outcome still depends on the target requirement. In JavaScript, four of the five requirements reach secure-and-correct outcomes in a single pass, and only R5 needs one additional iteration. Python shows a similar pattern: R2, R3, and R5 succeed immediately, while R1 and R4 converge after bounded iteration. PHP is less consistent. R1 and R2 succeed in a single pass, and R4 converges at iteration 4, but R3 and R5 do not reach secure-and-correct outcomes within the budget. Overall, narrowing the repair scope from the full project to an individual requirement improves effectiveness, but harder requirements can still fail to converge.

\begin{table}
\caption{RQ1: Pipeline-level effectiveness across languages under comparable budgets.}
\label{tab:rq1_pipeline_effectiveness}
\begin{tblr}{
  row{1} = {c},
  cell{2}{1} = {r=12}{c},
  cell{2}{5} = {c},
  cell{2}{6} = {c},
  cell{2}{7} = {c},
  cell{2}{8} = {c},
  cell{3}{5} = {c},
  cell{3}{6} = {c},
  cell{3}{7} = {c},
  cell{3}{8} = {c},
  cell{4}{5} = {c},
  cell{4}{6} = {c},
  cell{4}{7} = {c},
  cell{4}{8} = {c},
  cell{5}{5} = {c},
  cell{5}{6} = {c},
  cell{5}{7} = {c},
  cell{5}{8} = {c},
  cell{6}{5} = {c},
  cell{6}{6} = {c},
  cell{6}{7} = {c},
  cell{6}{8} = {c},
  cell{7}{5} = {c},
  cell{7}{6} = {c},
  cell{7}{7} = {c},
  cell{7}{8} = {c},
  cell{8}{5} = {c},
  cell{8}{6} = {c},
  cell{8}{7} = {c},
  cell{8}{8} = {c},
  cell{9}{5} = {c},
  cell{9}{6} = {c},
  cell{9}{7} = {c},
  cell{9}{8} = {c},
  cell{10}{5} = {c},
  cell{10}{6} = {c},
  cell{10}{7} = {c},
  cell{10}{8} = {c},
  cell{11}{5} = {c},
  cell{11}{6} = {c},
  cell{11}{7} = {c},
  cell{11}{8} = {c},
  cell{12}{5} = {c},
  cell{12}{6} = {c},
  cell{12}{7} = {c},
  cell{12}{8} = {c},
  cell{13}{3} = {font=\bfseries},
  cell{13}{4} = {font=\bfseries},
  cell{13}{5} = {c},
  cell{13}{6} = {c},
  cell{13}{7} = {c},
  cell{13}{8} = {c,font=\bfseries},
  cell{14}{1} = {r=20}{c},
  cell{14}{5} = {c},
  cell{14}{6} = {c},
  cell{14}{7} = {c},
  cell{14}{8} = {c,font=\bfseries},
  cell{15}{3} = {font=\bfseries},
  cell{15}{4} = {font=\bfseries},
  cell{15}{5} = {c},
  cell{15}{6} = {c},
  cell{15}{7} = {c},
  cell{15}{8} = {c,font=\bfseries},
  cell{16}{3} = {font=\bfseries},
  cell{16}{4} = {font=\bfseries},
  cell{16}{5} = {c},
  cell{16}{6} = {c},
  cell{16}{7} = {c},
  cell{16}{8} = {c,font=\bfseries},
  cell{17}{3} = {font=\bfseries},
  cell{17}{4} = {font=\bfseries},
  cell{17}{5} = {c},
  cell{17}{6} = {c},
  cell{17}{7} = {c},
  cell{17}{8} = {c,font=\bfseries},
  cell{18}{5} = {c},
  cell{18}{6} = {c},
  cell{18}{7} = {c},
  cell{18}{8} = {c},
  cell{19}{5} = {c},
  cell{19}{6} = {c},
  cell{19}{7} = {c},
  cell{19}{8} = {c},
  cell{20}{5} = {c},
  cell{20}{6} = {c},
  cell{20}{7} = {c},
  cell{20}{8} = {c},
  cell{21}{5} = {c},
  cell{21}{6} = {c},
  cell{21}{7} = {c},
  cell{21}{8} = {c},
  cell{22}{5} = {c},
  cell{22}{6} = {c},
  cell{22}{7} = {c},
  cell{22}{8} = {c,font=\bfseries},
  cell{23}{3} = {font=\bfseries},
  cell{23}{4} = {font=\bfseries},
  cell{23}{5} = {c},
  cell{23}{6} = {c},
  cell{23}{7} = {c},
  cell{23}{8} = {c,font=\bfseries},
  cell{24}{3} = {font=\bfseries},
  cell{24}{4} = {font=\bfseries},
  cell{24}{5} = {c},
  cell{24}{6} = {c},
  cell{24}{7} = {c},
  cell{24}{8} = {c,font=\bfseries},
  cell{25}{3} = {font=\bfseries},
  cell{25}{4} = {font=\bfseries},
  cell{25}{5} = {c},
  cell{25}{6} = {c},
  cell{25}{7} = {c},
  cell{25}{8} = {c,font=\bfseries},
  cell{26}{5} = {c},
  cell{26}{6} = {c},
  cell{26}{7} = {c},
  cell{26}{8} = {c,font=\bfseries},
  cell{27}{3} = {font=\bfseries},
  cell{27}{4} = {font=\bfseries},
  cell{27}{5} = {c},
  cell{27}{6} = {c},
  cell{27}{7} = {c},
  cell{27}{8} = {c,font=\bfseries},
  cell{28}{3} = {font=\bfseries},
  cell{28}{4} = {font=\bfseries},
  cell{28}{5} = {c},
  cell{28}{6} = {c},
  cell{28}{7} = {c},
  cell{28}{8} = {c,font=\bfseries},
  cell{29}{3} = {font=\bfseries},
  cell{29}{4} = {font=\bfseries},
  cell{29}{5} = {c},
  cell{29}{6} = {c},
  cell{29}{7} = {c},
  cell{29}{8} = {c,font=\bfseries},
  cell{30}{5} = {c},
  cell{30}{6} = {c},
  cell{30}{7} = {c},
  cell{30}{8} = {c,font=\bfseries},
  cell{31}{3} = {font=\bfseries},
  cell{31}{4} = {font=\bfseries},
  cell{31}{5} = {c},
  cell{31}{6} = {c},
  cell{31}{7} = {c},
  cell{31}{8} = {c,font=\bfseries},
  cell{32}{3} = {font=\bfseries},
  cell{32}{4} = {font=\bfseries},
  cell{32}{5} = {c},
  cell{32}{6} = {c},
  cell{32}{7} = {c},
  cell{32}{8} = {c,font=\bfseries},
  cell{33}{3} = {font=\bfseries},
  cell{33}{4} = {font=\bfseries},
  cell{33}{5} = {c},
  cell{33}{6} = {c},
  cell{33}{7} = {c},
  cell{33}{8} = {c,font=\bfseries},
  cell{34}{1} = {r=15}{c},
  cell{34}{5} = {c},
  cell{34}{6} = {c},
  cell{34}{7} = {c},
  cell{34}{8} = {c,font=\bfseries},
  cell{35}{3} = {font=\bfseries},
  cell{35}{4} = {font=\bfseries},
  cell{35}{5} = {c},
  cell{35}{6} = {c},
  cell{35}{7} = {c},
  cell{35}{8} = {c,font=\bfseries},
  cell{36}{3} = {font=\bfseries},
  cell{36}{4} = {font=\bfseries},
  cell{36}{5} = {c},
  cell{36}{6} = {c},
  cell{36}{7} = {c},
  cell{36}{8} = {c,font=\bfseries},
  cell{37}{5} = {c},
  cell{37}{6} = {c},
  cell{37}{7} = {c},
  cell{37}{8} = {c,font=\bfseries},
  cell{38}{3} = {font=\bfseries},
  cell{38}{4} = {font=\bfseries},
  cell{38}{5} = {c},
  cell{38}{6} = {c},
  cell{38}{7} = {c},
  cell{38}{8} = {c,font=\bfseries},
  cell{39}{5} = {c},
  cell{39}{6} = {c},
  cell{39}{7} = {c},
  cell{39}{8} = {c},
  cell{40}{5} = {c},
  cell{40}{6} = {c},
  cell{40}{7} = {c},
  cell{40}{8} = {c},
  cell{41}{5} = {c},
  cell{41}{6} = {c},
  cell{41}{7} = {c},
  cell{41}{8} = {c},
  cell{42}{5} = {c},
  cell{42}{6} = {c},
  cell{42}{7} = {c},
  cell{42}{8} = {c},
  cell{43}{5} = {c},
  cell{43}{6} = {c},
  cell{43}{7} = {c},
  cell{43}{8} = {c,font=\bfseries},
  cell{44}{3} = {font=\bfseries},
  cell{44}{4} = {font=\bfseries},
  cell{44}{5} = {c},
  cell{44}{6} = {c},
  cell{44}{7} = {c},
  cell{44}{8} = {c,font=\bfseries},
  cell{45}{3} = {font=\bfseries},
  cell{45}{4} = {font=\bfseries},
  cell{45}{5} = {c},
  cell{45}{6} = {c},
  cell{45}{7} = {c},
  cell{45}{8} = {c,font=\bfseries},
  cell{46}{3} = {font=\bfseries},
  cell{46}{4} = {font=\bfseries},
  cell{46}{5} = {c},
  cell{46}{6} = {c},
  cell{46}{7} = {c},
  cell{46}{8} = {c,font=\bfseries},
  cell{47}{5} = {c},
  cell{47}{6} = {c},
  cell{47}{7} = {c},
  cell{47}{8} = {c},
  cell{48}{5} = {c},
  cell{48}{6} = {c},
  cell{48}{7} = {c},
  cell{48}{8} = {c},
  hline{1-2,14,34,49} = {-}{0.08em},
  hline{4,6,8,10,12,18,20,22,26,30,37,39,41,43,47} = {2-8}{dotted},
}
Language   & Granularity       & Req. & Workflow       & {Used~\\Iterations} & {Func.\\~Pass Rate} & {Exploit~\\Success~\\Rate} & {S\textbackslash{}C~\\Yield Rate} \\
JavaScript & Project-level     & R1–R5       & Single-pass DRV& 1                   & 1.0                 & 0.0                      & 1.0                               \\
           & Project-level     & R1–R5       & Iterative DRV  & –                   & –                   & –                        & –                                 \\
           & Requirement-level & R1          & Single-pass DRV & 1                   & 1.0                 & 0.0                      & 1.0                               \\
           & Requirement-level & R1          & Iterative DRV  & –                   & –                   & –                        & –                                 \\
           & Requirement-level & R2          & Single-pass DRV & 1                   & 1.0                 & 0.0                      & 1.0                               \\
           & Requirement-level & R2          & Iterative DRV  & –                   & –                   & –                        & –                                 \\
           & Requirement-level & R3          & Single-pass DRV & 1                   & 1.0                 & 0.0                      & 1.0                               \\
           & Requirement-level & R3          & Iterative DRV  & –                   & –                   & –                        & –                                 \\
           & Requirement-level & R4          & Single-pass DRV & 1                   & 1.0                 & 0.0                      & 1.0                               \\
           & Requirement-level & R4          & Iterative DRV  & –                   & –                   & –                        & –                                 \\
           & Requirement-level & R5          & Single-pass DRV & 1                   & 1.0                 & 1.0                      & 0.0                               \\
           & Requirement-level & R5          & Iterative DRV  & 2                   & 1.0                 & 0.0                      & 1.0                               \\
PHP        & Project-level     & R1–R5       & Single-pass DRV & 1                   & 0.0                 & 1.0                      & 0.0                               \\
           & Project-level     & R1–R5       & Iterative DRV  & 2                   & 0.0                 & 0.0                      & 0.0                               \\
           & Project-level     & R1–R5       & Iterative DRV  & 3                   & 0.0                 & 0.0                      & 0.0                               \\
           & Project-level     & R1–R5       & Iterative DRV  & 4                   & 0.0                 & 0.0                      & 0.0                               \\
           & Requirement-level & R1          & Single-pass DRV & 1                   & 1.0                 & 0.0                      & 1.0                               \\
           & Requirement-level & R1          & Iterative DRV  & –                   & –                   & –                        & –                                 \\
           & Requirement-level & R2          & Single-pass DRV & 1                   & 1.0                 & 0.0                      & 1.0                               \\
           & Requirement-level & R2          & Iterative DRV  & –                   & –                   & –                        & –                                 \\
           & Requirement-level & R3          & Single-pass DRV & 1                   & 0.0                 & 0.0                      & 0.0                               \\
           & Requirement-level & R3          & Iterative DRV  & 2                   & 0.0                 & 0.0                      & 0.0                               \\
           & Requirement-level & R3          & Iterative DRV  & 3                   & 0.66                & 0.0                      & 0.0                               \\
           & Requirement-level & R3          & Iterative DRV  & 4                   & 0.66                & 0.0                      & 0.0                               \\
           & Requirement-level & R4          & Single-pass DRV & 1                   & 0.5                 & 0.0                      & 0.0                               \\
           & Requirement-level & R4          & Iterative DRV  & 2                   & 0.5                 & 0.0                      & 0.0                               \\
           & Requirement-level & R4          & Iterative DRV  & 3                   & 0.5                 & 0.0                      & 0.0                               \\
           & Requirement-level & R4          & Iterative DRV  & 4                   & 1.0                 & 0.0                      & 1.0                               \\
           & Requirement-level & R5          & Single-pass DRV & 1                   & 0.0                 & 0.0                      & 0.0                               \\
           & Requirement-level & R5          & Iterative DRV  & 2                   & 0.0                 & 0.0                      & 0.0                               \\
           & Requirement-level & R5          & Iterative DRV  & 3                   & 0.0                 & 0.0                      & 0.0                               \\
           & Requirement-level & R5          & Iterative DRV  & 4                   & 0.0                 & 0.0                      & 0.0                               \\
Python     & Project-level     & R1–R5       & Single-pass DRV & 1                   & 0.0                 & 1.0                      & 0.0                               \\
           & Project-level     & R1–R5       & Iterative DRV  & 2                   & 1.0                 & 0.0(errors)              & 0.0                               \\
           & Project-level     & R1–R5       & Iterative DRV  & 3                   & 1.0                 & 0.0                      & 1.0                               \\
           & Requirement-level & R1          & Single-pass DRV & 1                   & 1.0                 & 0.25                     & 0.0                               \\
           & Requirement-level  & R1          & Iterative DRV  & 2                   & 1.0                 & 0.00                     & 1.0                               \\
           & Requirement-level & R2          & Single-pass DRV & 1                   & 1.0                 & 0.0                      & 1.0                               \\
           & Requirement-level & R2          & Iterative DRV  & –                   & –                   & –                        & –                                 \\
           & Requirement-level & R3          & Single-pass DRV & 1                   & 1.0                 & 0.0                      & 1.0                               \\
           & Requirement-level & R3          & Iterative DRV  & –                   & –                   & –                        & –                                 \\
           & Requirement-level & R4          & Single-pass DRV & 1                   & 0.5                 & 0.5                      & 0.0                               \\
           & Requirement-level & R4          & Iterative DRV  & 2                   & 0.5                 & 0.0                      & 0.0                               \\
           & Requirement-level & R4          & Iterative DRV  & 3                   & 0.5                 & 0.0                      & 0.0                               \\
           & Requirement-level & R4          & Iterative DRV  & 4                   & 1.0                 & 0.0                      & 1.0                               \\
           & Requirement-level & R5          & Single-pass DRV & 1                   & 1.0                 & 0.0                      & 1.0                               \\
           & Requirement-level & R5          & Iterative DRV  & –                   & –                   & –                        & –                                 
\end{tblr}
\end{table}

\paragraph{File-level observations.}
The file-level results show the strongest pipeline effectiveness across the evaluated settings. As shown in Tables~\ref{tab:file_level_result_chatgpt} and~\ref{tab:file_level_result_glm}, most files reach secure-and-correct outcomes in a single pass for both ChatGPT-5 and GLM-5. For ChatGPT-5, only a small number of difficult cases require bounded iteration, most notably JavaScript Vul002 and Vul005, as well as Python Vul005 (Table~\ref{tab:file_level_result_chatgpt}). For GLM-5, file-level performance is similarly strong: most files succeed in a single pass, and only a few JavaScript cases require iterative refinement, notably Vul002 and Vul005 (Table~\ref{tab:file_level_result_glm}). PHP reaches secure-and-correct outcomes in a single pass for all file-level cases under both models (Tables~\ref{tab:file_level_result_chatgpt} and~\ref{tab:file_level_result_glm}). Taken together, these results show a clear granularity effect: as the repair target becomes more localized, D--R--V becomes more effective and secure-and-correct convergence becomes more consistent.

\begin{table}
\centering
\caption{ChatGPT-5---File-level experimental record for single-pass and iterative workflows.}
\label{tab:file_level_result_chatgpt}
\begin{tblr}{
  row{1} = {c},
  cell{2}{1} = {r=13}{},
  cell{2}{2} = {r=13}{},
  cell{2}{5} = {c},
  cell{2}{6} = {c},
  cell{2}{7} = {c},
  cell{2}{8} = {c},
  cell{3}{5} = {c},
  cell{3}{6} = {c},
  cell{3}{7} = {c},
  cell{3}{8} = {c},
  cell{4}{5} = {c},
  cell{4}{6} = {c},
  cell{4}{7} = {c},
  cell{4}{8} = {c},
  cell{5}{3} = {font=\bfseries},
  cell{5}{4} = {font=\bfseries},
  cell{5}{5} = {c},
  cell{5}{6} = {c},
  cell{5}{7} = {c},
  cell{5}{8} = {c,font=\bfseries},
  cell{6}{3} = {font=\bfseries},
  cell{6}{4} = {font=\bfseries},
  cell{6}{5} = {c},
  cell{6}{6} = {c},
  cell{6}{7} = {c},
  cell{6}{8} = {c,font=\bfseries},
  cell{7}{5} = {c},
  cell{7}{6} = {c},
  cell{7}{7} = {c},
  cell{7}{8} = {c},
  cell{8}{5} = {c},
  cell{8}{6} = {c},
  cell{8}{7} = {c},
  cell{8}{8} = {c},
  cell{9}{5} = {c},
  cell{9}{6} = {c},
  cell{9}{7} = {c},
  cell{9}{8} = {c},
  cell{10}{5} = {c},
  cell{10}{6} = {c},
  cell{10}{7} = {c},
  cell{10}{8} = {c},
  cell{11}{5} = {c},
  cell{11}{6} = {c},
  cell{11}{7} = {c},
  cell{11}{8} = {c,font=\bfseries},
  cell{12}{3} = {font=\bfseries},
  cell{12}{4} = {font=\bfseries},
  cell{12}{5} = {c},
  cell{12}{6} = {c},
  cell{12}{7} = {c},
  cell{12}{8} = {c,font=\bfseries},
  cell{13}{5} = {c},
  cell{13}{6} = {c},
  cell{13}{7} = {c},
  cell{13}{8} = {c},
  cell{14}{5} = {c},
  cell{14}{6} = {c},
  cell{14}{7} = {c},
  cell{14}{8} = {c},
  cell{15}{1} = {r=14}{},
  cell{15}{2} = {r=14}{},
  cell{15}{5} = {c},
  cell{15}{6} = {c},
  cell{15}{7} = {c},
  cell{15}{8} = {c,font=\bfseries},
  cell{16}{5} = {c},
  cell{16}{6} = {c},
  cell{16}{7} = {c},
  cell{16}{8} = {c},
  cell{17}{5} = {c},
  cell{17}{6} = {c},
  cell{17}{7} = {c},
  cell{17}{8} = {c},
  cell{18}{5} = {c},
  cell{18}{6} = {c},
  cell{18}{7} = {c},
  cell{18}{8} = {c},
  cell{19}{5} = {c},
  cell{19}{6} = {c},
  cell{19}{7} = {c},
  cell{19}{8} = {c},
  cell{20}{5} = {c},
  cell{20}{6} = {c},
  cell{20}{7} = {c},
  cell{20}{8} = {c},
  cell{21}{5} = {c},
  cell{21}{6} = {c},
  cell{21}{7} = {c},
  cell{21}{8} = {c},
  cell{22}{5} = {c},
  cell{22}{6} = {c},
  cell{22}{7} = {c},
  cell{22}{8} = {c},
  cell{23}{5} = {c},
  cell{23}{6} = {c},
  cell{23}{7} = {c},
  cell{23}{8} = {c,font=\bfseries},
  cell{24}{3} = {font=\bfseries},
  cell{24}{4} = {font=\bfseries},
  cell{24}{5} = {c},
  cell{24}{6} = {c},
  cell{24}{7} = {c},
  cell{24}{8} = {c,font=\bfseries},
  cell{25}{5} = {c},
  cell{25}{6} = {c},
  cell{25}{7} = {c},
  cell{25}{8} = {c},
  cell{26}{5} = {c},
  cell{26}{6} = {c},
  cell{26}{7} = {c},
  cell{26}{8} = {c},
  cell{27}{5} = {c},
  cell{27}{6} = {c},
  cell{27}{7} = {c},
  cell{27}{8} = {c},
  cell{28}{5} = {c},
  cell{28}{6} = {c},
  cell{28}{7} = {c},
  cell{28}{8} = {c},
  cell{29}{1} = {r=14}{},
  cell{29}{2} = {r=14}{},
  cell{29}{5} = {c},
  cell{29}{6} = {c},
  cell{29}{7} = {c},
  cell{29}{8} = {c},
  cell{30}{5} = {c},
  cell{30}{6} = {c},
  cell{30}{7} = {c},
  cell{30}{8} = {c},
  cell{31}{5} = {c},
  cell{31}{6} = {c},
  cell{31}{7} = {c},
  cell{31}{8} = {c},
  cell{32}{5} = {c},
  cell{32}{6} = {c},
  cell{32}{7} = {c},
  cell{32}{8} = {c},
  cell{33}{5} = {c},
  cell{33}{6} = {c},
  cell{33}{7} = {c},
  cell{33}{8} = {c},
  cell{34}{5} = {c},
  cell{34}{6} = {c},
  cell{34}{7} = {c},
  cell{34}{8} = {c},
  cell{35}{5} = {c},
  cell{35}{6} = {c},
  cell{35}{7} = {c},
  cell{35}{8} = {c},
  cell{36}{5} = {c},
  cell{36}{6} = {c},
  cell{36}{7} = {c},
  cell{36}{8} = {c},
  cell{37}{5} = {c},
  cell{37}{6} = {c},
  cell{37}{7} = {c},
  cell{37}{8} = {c},
  cell{38}{5} = {c},
  cell{38}{6} = {c},
  cell{38}{7} = {c},
  cell{38}{8} = {c},
  cell{39}{5} = {c},
  cell{39}{6} = {c},
  cell{39}{7} = {c},
  cell{39}{8} = {c},
  cell{40}{5} = {c},
  cell{40}{6} = {c},
  cell{40}{7} = {c},
  cell{40}{8} = {c},
  cell{41}{5} = {c},
  cell{41}{6} = {c},
  cell{41}{7} = {c},
  cell{41}{8} = {c},
  cell{42}{5} = {c},
  cell{42}{6} = {c},
  cell{42}{7} = {c},
  cell{42}{8} = {c},
  hline{1-2} = {-}{},
  hline{4,7,9,11,13,17,19,21,23,25,27,31,33,35,37,39,41} = {3-8}{dotted},
  hline{15,29,43} = {-}{0.08em},
}
Language   & Granularity & File   & Workflow       & {Used~\\Iterations} & {Func. Pass~\\Rate} & {Exploit~\\Success~\\Rate} & {S\textbackslash{}C Yield~\\Rate} \\
JavaScript & File-level  & Vul001 & Single-pass DRV& 1                   & 1.0                 & 0.0                      & 1.0                               \\
           &             & Vul001 & Iterative DRV  & –                   & –                   & –                        & –                                 \\
           &             & Vul002 & Single-pass DRV & 1                   & 1.0                 & 1.0                      & 0.0                               \\
           &             & Vul002 & Iterative DRV  & 2                   & 1.0                 & 1.0                      & 0.0                               \\
           &             & Vul002 & Iterative DRV  & 3                   & 1.0                 & 0.0                      & 1.0                               \\
           &             & Vul003 & Single-pass DRV & 1                   & 1.0                 & 0.0                      & 1.0                               \\
           &             & Vul003 & Iterative DRV  & –                   & –                   & –                        & –                                 \\
           &             & Vul004 & Single-pass DRV & 1                   & 1.0                 & 0.0                      & 1.0                               \\
           &             & Vul004 & Iterative DRV  & –                   & –                   & –                        & –                                 \\
           &             & Vul005 & Single-pass DRV & 1                   & 1.0                 & 1.0                      & 0.0                               \\
           &             & Vul005 & Iterative DRV  & 2                   & 1.0                 & 0.0                      & 1.0                               \\
           &             & Vul006 & Single-pass DRV & 1                   & 1.0                 & 0.0                      & 1.0                               \\
           &             & Vul006 & Iterative DRV  & –                   & –                   & –                        & –                                 \\
Python     & File-level  & Vul001 & Single-pass DRV & 1                   & 1.0                 & 0.0                      & 1.0                               \\
           &             & Vul001 & Iterative DRV  & –                   & –                   & –                        & –                                 \\
           &             & Vul002 & Single-pass DRV & 1                   & 1.0                 & 0.0                      & 1.0                               \\
           &             & Vul002 & Iterative DRV  & –                   & –                   & –                        & –                                 \\
           &             & Vul003 & Single-pass DRV & 1                   & 1.0                 & 0.0                      & 1.0                               \\
           &             & Vul003 & Iterative DRV  & –                   & –                   & –                        & –                                 \\
           &             & Vul004 & Single-pass DRV & 1                   & 1.0                 & 0.0                      & 1.0                               \\
           &             & Vul004 & Iterative DRV  & –                   & –                   & –                        & –                                 \\
           &             & Vul005 & Single-pass DRV & 1                   & 0.8                 & 0.0                      & 0.0                               \\
           &             & Vul005 & Iterative DRV  & 2                   & 1.0                 & 0.0                      & 1.0                               \\
           &             & Vul006 & Single-pass DRV & 1                   & 1.0                 & 0.0                      & 1.0                               \\
           &             & Vul006 & Iterative DRV  & –                   & –                   & –                        & –                                 \\
           &             & Vul007 & Single-pass DRV & 1                   & 1.0                 & 0.0                      & 1.0                               \\
           &             & Vul007 & Iterative DRV  & –                   & –                   & –                        & –                                 \\
PHP        & File-level  & Vul001 & Single-pass DRV & 1                   & 1.0                 & 0.0                      & 1.0                               \\
           &             & Vul001 & Iterative DRV  & –                   & –                   & –                        & –                                 \\
           &             & Vul002 & Single-pass DRV & 1                   & 1.0                 & 0.0                      & 1.0                               \\
           &             & Vul002 & Iterative DRV  & –                   & –                   & –                        & –                                 \\
           &             & Vul003 & Single-pass DRV & 1                   & 1.0                 & 0.0                      & 1.0                               \\
           &             & Vul003 & Iterative DRV  & –                   & –                   & –                        & –                                 \\
           &             & Vul004 & Single-pass DRV & 1                   & 1.0                 & 0.0                      & 1.0                               \\
           &             & Vul004 & Iterative DRV  & –                   & –                   & –                        & –                                 \\
           &             & Vul005 & Single-pass DRV & 1                   & 1.0                 & 0.0                      & 1.0                               \\
           &             & Vul005 & Iterative DRV  & –                   & –                   & –                        & –                                 \\
           &             & Vul006 & Single-pass DRV & 1                   & 1.0                 & 0.0                      & 1.0                               \\
           &             & Vul006 & Iterative DRV  & –                   & –                   & –                        & –                                 \\
           &             & Vul007 & Single-pass DRV & 1                   & 1.0                 & 0.0                      & 1.0                               \\
           &             & Vul007 & Iterative DRV  & –                   & –                   & –                        & –                                 
\end{tblr}
\end{table}

\begin{table}
\centering
\caption{GLM-5---File-level experimental record for single-pass and iterative workflows.}
\label{tab:file_level_result_glm}
\begin{tblr}{
  row{1} = {c},
  cell{2}{1} = {r=12}{},
  cell{2}{2} = {r=12}{},
  cell{2}{5} = {c},
  cell{2}{6} = {c},
  cell{2}{7} = {c},
  cell{2}{8} = {c},
  cell{3}{5} = {c},
  cell{3}{6} = {c},
  cell{3}{7} = {c},
  cell{3}{8} = {c},
  cell{4}{5} = {c},
  cell{4}{6} = {c},
  cell{4}{7} = {c},
  cell{4}{8} = {c,font=\bfseries},
  cell{5}{3} = {font=\bfseries},
  cell{5}{4} = {font=\bfseries},
  cell{5}{5} = {c},
  cell{5}{6} = {c},
  cell{5}{7} = {c},
  cell{5}{8} = {c,font=\bfseries},
  cell{6}{5} = {c},
  cell{6}{6} = {c},
  cell{6}{7} = {c},
  cell{6}{8} = {c},
  cell{7}{5} = {c},
  cell{7}{6} = {c},
  cell{7}{7} = {c},
  cell{7}{8} = {c},
  cell{8}{5} = {c},
  cell{8}{6} = {c},
  cell{8}{7} = {c},
  cell{8}{8} = {c},
  cell{9}{5} = {c},
  cell{9}{6} = {c},
  cell{9}{7} = {c},
  cell{9}{8} = {c},
  cell{10}{5} = {c},
  cell{10}{6} = {c},
  cell{10}{7} = {c},
  cell{10}{8} = {c,font=\bfseries},
  cell{11}{3} = {font=\bfseries},
  cell{11}{4} = {font=\bfseries},
  cell{11}{5} = {c},
  cell{11}{6} = {c},
  cell{11}{7} = {c},
  cell{11}{8} = {c,font=\bfseries},
  cell{12}{5} = {c},
  cell{12}{6} = {c},
  cell{12}{7} = {c},
  cell{12}{8} = {c},
  cell{13}{5} = {c},
  cell{13}{6} = {c},
  cell{13}{7} = {c},
  cell{13}{8} = {c},
  cell{14}{1} = {r=14}{},
  cell{14}{2} = {r=14}{},
  cell{14}{5} = {c},
  cell{14}{6} = {c},
  cell{14}{7} = {c},
  cell{14}{8} = {c},
  cell{15}{5} = {c},
  cell{15}{6} = {c},
  cell{15}{7} = {c},
  cell{15}{8} = {c},
  cell{16}{5} = {c},
  cell{16}{6} = {c},
  cell{16}{7} = {c},
  cell{16}{8} = {c},
  cell{17}{5} = {c},
  cell{17}{6} = {c},
  cell{17}{7} = {c},
  cell{17}{8} = {c},
  cell{18}{5} = {c},
  cell{18}{6} = {c},
  cell{18}{7} = {c},
  cell{18}{8} = {c},
  cell{19}{5} = {c},
  cell{19}{6} = {c},
  cell{19}{7} = {c},
  cell{19}{8} = {c},
  cell{20}{5} = {c},
  cell{20}{6} = {c},
  cell{20}{7} = {c},
  cell{20}{8} = {c},
  cell{21}{5} = {c},
  cell{21}{6} = {c},
  cell{21}{7} = {c},
  cell{21}{8} = {c},
  cell{22}{5} = {c},
  cell{22}{6} = {c},
  cell{22}{7} = {c},
  cell{22}{8} = {c},
  cell{23}{5} = {c},
  cell{23}{6} = {c},
  cell{23}{7} = {c},
  cell{23}{8} = {c},
  cell{24}{5} = {c},
  cell{24}{6} = {c},
  cell{24}{7} = {c},
  cell{24}{8} = {c},
  cell{25}{5} = {c},
  cell{25}{6} = {c},
  cell{25}{7} = {c},
  cell{25}{8} = {c},
  cell{26}{5} = {c},
  cell{26}{6} = {c},
  cell{26}{7} = {c},
  cell{26}{8} = {c},
  cell{27}{5} = {c},
  cell{27}{6} = {c},
  cell{27}{7} = {c},
  cell{27}{8} = {c},
  cell{28}{1} = {r=14}{},
  cell{28}{2} = {r=14}{},
  cell{28}{5} = {c},
  cell{28}{6} = {c},
  cell{28}{7} = {c},
  cell{28}{8} = {c},
  cell{29}{5} = {c},
  cell{29}{6} = {c},
  cell{29}{7} = {c},
  cell{29}{8} = {c},
  cell{30}{5} = {c},
  cell{30}{6} = {c},
  cell{30}{7} = {c},
  cell{30}{8} = {c},
  cell{31}{5} = {c},
  cell{31}{6} = {c},
  cell{31}{7} = {c},
  cell{31}{8} = {c},
  cell{32}{5} = {c},
  cell{32}{6} = {c},
  cell{32}{7} = {c},
  cell{32}{8} = {c},
  cell{33}{5} = {c},
  cell{33}{6} = {c},
  cell{33}{7} = {c},
  cell{33}{8} = {c},
  cell{34}{5} = {c},
  cell{34}{6} = {c},
  cell{34}{7} = {c},
  cell{34}{8} = {c},
  cell{35}{5} = {c},
  cell{35}{6} = {c},
  cell{35}{7} = {c},
  cell{35}{8} = {c},
  cell{36}{5} = {c},
  cell{36}{6} = {c},
  cell{36}{7} = {c},
  cell{36}{8} = {c},
  cell{37}{5} = {c},
  cell{37}{6} = {c},
  cell{37}{7} = {c},
  cell{37}{8} = {c},
  cell{38}{5} = {c},
  cell{38}{6} = {c},
  cell{38}{7} = {c},
  cell{38}{8} = {c},
  cell{39}{5} = {c},
  cell{39}{6} = {c},
  cell{39}{7} = {c},
  cell{39}{8} = {c},
  cell{40}{5} = {c},
  cell{40}{6} = {c},
  cell{40}{7} = {c},
  cell{40}{8} = {c},
  cell{41}{5} = {c},
  cell{41}{6} = {c},
  cell{41}{7} = {c},
  cell{41}{8} = {c},
  hline{1-2,14,28,42} = {-}{},
  hline{4,6,8,10,12,16,18,20,22,24,26,30,32,34,36,38,40} = {3-8}{dotted},
}
Language   & Granularity & File   & Workflow       & {Used~\\Iterations} & {Func. Pass~\\Rate} & {Exploit~\\Success~\\Rate} & {S\textbackslash{}C Yield~\\Rate} \\
JavaScript & File-level  & Vul001 & Single-pass DRV& 1                   & 1.0                 & 0.0                & 1.0                               \\
           &             & Vul001 & Iterative DRV  & –                   & –                   & –                  & –                                 \\
           &             & Vul002 & Single-pass DRV & 1                   & 1.0                 & 0.0                & 0.0                               \\
           &             & Vul002 & Iterative DRV  & 2                   & 1.0                 & 0.0                & 1.0                               \\
           &             & Vul003 & Single-pass DRV & 1                   & 1.0                 & 0.0                & 1.0                               \\
           &             & Vul003 & Iterative DRV  & –                   & –                   & –                  & –                                 \\
           &             & Vul004 & Single-pass DRV & 1                   & 1.0                 & 0.0                & 1.0                               \\
           &             & Vul004 & Iterative DRV  & –                   & –                   & –                  & –                                 \\
           &             & Vul005 & Single-pass DRV & 1                   & 1.0                 & 1.0                & 0.0                               \\
           &             & Vul005 & Iterative DRV  & 2                   & 1.0                 & 0.0                & 1.0                               \\
           &             & Vul006 & Single-pass DRV & 1                   & 1.0                 & 0.0                & 1.0                               \\
           &             & Vul006 & Iterative DRV  & 2                   & –                   & –                  & –                                 \\
Python     & File-level  & Vul001 & Single-pass DRV & 1                   & 1.0                 & 0.0                & 1.0                               \\
           &             & Vul001 & Iterative DRV  & –                   & –                   & –                  & –                                 \\
           &             & Vul002 & Single-pass DRV & 1                   & 1.0                 & 0.0                & 1.0                               \\
           &             & Vul002 & Iterative DRV  & –                   & –                   & –                  & –                                 \\
           &             & Vul003 & Single-pass DRV & 1                   & 1.0                 & 0.0                & 1.0                               \\
           &             & Vul003 & Iterative DRV  & –                   & –                   & –                  & –                                 \\
           &             & Vul004 & Single-pass DRV & 1                   & 1.0                 & 0.0                & 1.0                               \\
           &             & Vul004 & Iterative DRV  & –                   & –                   & –                  & –                                 \\
           &             & Vul005 & Single-pass DRV & 1                   & 1.0                 & 0.0                & 1.0                               \\
           &             & Vul005 & Iterative DRV  & –                   & –                   & –                  & –                                 \\
           &             & Vul006 & Single-pass DRV & 1                   & 1.0                 & 0.0                & 1.0                               \\
           &             & Vul006 & Iterative DRV  & –                   & –                   & –                  & –                                 \\
           &             & Vul007 & Single-pass DRV & 1                   & 1.0                 & 0.0                & 1.0                               \\
           &             & Vul007 & Iterative DRV  & –                   & –                   & –                  & –                                 \\
PHP        & File-level  & Vul001 & Single-pass DRV & 1                   & 1.0                 & 0.0                & 1.0                               \\
           &             & Vul001 & Iterative DRV  & –                   & –                   & –                  & –                                 \\
           &             & Vul002 & Single-pass DRV & 1                   & 1.0                 & 0.0                & 1.0                               \\
           &             & Vul002 & Iterative DRV  & –                   & –                   & –                  & –                                 \\
           &             & Vul003 & Single-pass DRV & 1                   & 1.0                 & 0.0                & 1.0                               \\
           &             & Vul003 & Iterative DRV  & –                   & –                   & –                  & –                                 \\
           &             & Vul004 & Single-pass DRV& 1                   & 1.0                 & 0.0                & 1.0                               \\
           &             & Vul004 & Iterative DRV  & –                   & –                   & –                  & –                                 \\
           &             & Vul005 & Single-pass DRV & 1                   & 1.0                 & 0.0                & 1.0                               \\
           &             & Vul005 & Iterative DRV  & –                   & –                   & –                  & –                                 \\
           &             & Vul006 & Single-pass DRV & 1                   & 1.0                 & 0.0                & 1.0                               \\
           &             & Vul006 & Iterative DRV  & –                   & –                   & –                  & –                                 \\
           &             & Vul007 & Single-pass DRV& 1                   & 1.0                 & 0.0                & 1.0                               \\
           &             & Vul007 & Iterative DRV  & –                   & –                   & –                  & –                                 
\end{tblr}
\end{table}

\begin{tcolorbox}[
  colback=gray!10,
  colframe=gray!60,
  boxrule=0.6pt,
  arc=2pt,
  left=6pt,right=6pt,top=4pt,bottom=4pt
]
\noindent \textbf{RQ1:} Bounded iterative detect--repair--verify workflows improve secure-and-correct yield over single-pass repair in some settings, but the gain is not uniform at the project level. The benefit becomes clearer and more consistent as the repair target narrows from the full project to individual requirements and files.
\end{tcolorbox}

\subsection{RQ2 (Detection reliability as repair guidance).}

\subsubsection{Measures}
RQ2 examines detection reports from two perspectives: \emph{repair actionability} and \emph{verification-grounded reliability}. The first asks whether a report gives enough concrete guidance to support downstream repair. The second asks whether that guidance is borne out by verified repair outcomes under the same workflow setting.

\paragraph{Repair actionability.}
For the project-level and requirement-level artifacts, repair actionability is assessed through the structural properties of the generated detection reports. Tables~\ref{tab:project_level_vd_guidance} and~\ref{tab:requirement_level_vd_guidance} summarize four indicators: \emph{\#VD Reports}, \emph{VD Structural Completeness}, \emph{Avg.\ Entry Points}, \emph{Avg.\ Evidence Files}, and \emph{Avg.\ Patch Locations}. Together, these indicators show whether a report identifies the relevant attack surface, grounds its claims in code-level evidence, and narrows the candidate repair region clearly enough to guide patching.

\paragraph{Verification-grounded reliability.}
Reliability is assessed by relating report structure to verified downstream repair outcomes. As shown in Tables~\ref{tab:project_level_verified_outcome} and~\ref{tab:requirement_level_verified_outcome_summary}, a report is treated as more reliable when its guidance supports a repair that reaches a secure-and-correct outcome under verification. In this sense, reliability is not determined by report completeness alone, but by whether the reported guidance is strong enough to support a patch that removes exploitability without breaking functionality.

\subsubsection{Findings.}

\paragraph{Project-level observations.} Project-level detection reports are complete in form, but their practical repair value differs across languages.
As shown in Table~\ref{tab:project_level_vd_guidance}, for the three project-level artifacts, all VD reports include the key elements needed for downstream repair, including entry points, code evidence, and candidate patch locations. This shows that the reports are not merely high-level vulnerability descriptions, but are designed to support repair. However, Table~\ref{tab:project_level_verified_outcome} shows that complete report structure does not by itself lead to successful repair. In JavaScript, the reported guidance is sufficient for a secure-and-correct result in a single pass. In Python, the same general pattern holds, but only after bounded iteration. In PHP, repair remains unsuccessful under verification despite similarly complete report structure. Taken together, Tables~\ref{tab:project_level_vd_guidance} and~\ref{tab:project_level_verified_outcome} suggest that project-level VD reports can provide useful repair guidance, but their real value depends on whether the repair stage can turn that guidance into a patch that preserves functionality and removes exploitability.

\begin{table}
\centering
\caption{Project-level characteristics of detection guidance.}
\label{tab:project_level_vd_guidance}
\begin{tblr}{
  row{1} = {c},
  cell{2}{3} = {c},
  cell{2}{4} = {c},
  cell{2}{5} = {c},
  cell{2}{6} = {c},
  cell{2}{7} = {c},
  cell{3}{3} = {c},
  cell{3}{4} = {c},
  cell{3}{5} = {c},
  cell{3}{6} = {c},
  cell{3}{7} = {c},
  cell{4}{3} = {c},
  cell{4}{4} = {c},
  cell{4}{5} = {c},
  cell{4}{6} = {c},
  cell{4}{7} = {c},
  hlines,
  hline{3-4} = {-}{dotted},
}
Language   & Artifact                 & {\#VD\\Reports} & {VD~\\Structural\\Completeness} & {Avg.~\\Entry\\Points} & {Avg.~\\Evidence\\Files} & {Avg.~\\Patch\\Locations} \\
JavaScript & {Project-level\\(R1--R5)} & 7               & 100\%                           & 3.00                   & 2.86                     & 2.14                      \\
PHP        & {Project-level\\(R1--R5)} & 8               & 100\%                           & 3.75                   & 2.62                     & 3.25                      \\
Python     & {Project-level\\(R1--R5)} & 9               & 100\%                           & 3.22                   & 2.00                     & 2.33                      
\end{tblr}
\end{table}

\begin{table}
\centering
\caption{Project-level verified repair outcomes under detection guidance.}
\label{tab:project_level_verified_outcome}
\begin{tblr}{
  row{1} = {c},
  hlines,
  hline{3-4} = {-}{dotted},
}
Language   & Verified Workflow Outcome                                                                                                  & Initial Interpretation                                                                                             \\
JavaScript & {Single-pass DRVachieves\\~Func.=1.0, Exploit=0.0, SC=1.0}                                                                 & {Reports appear highly~\\actionable and align\\~with a successful~\\single-pass repair outcome.}                   \\
PHP        & {Single-pass DRVremains at~\\Func.=0.0, Exploit=1.0, SC=0.0;~\\iterative DRV remains~\\unsuccessful through\\~iteration 4} & {Reports are structurally rich,~\\but verified repair remains\\~untrustworthy despite broad\\~claimed mitigation.} \\
Python     & {Single-pass DRVfails (0.0/1.0/0.0),~\\while iteration 3 reaches~\\Func.=1.0, Exploit=0.0, SC=1.0}                         & {Reports appear actionable,\\~but trustworthy repair is\\~established only after\\~bounded iteration.}             
\end{tblr}
\end{table}

\paragraph{Requirement-level observations.}
Requirement-level detection reports are complete in form, but their practical repair value still differs across requirements and languages. As shown in Table~\ref{tab:requirement_level_vd_guidance}, the available requirement-level VD reports include the key elements needed for downstream repair, including entry points, code evidence, and candidate patch locations. This shows that the reports are not merely high-level vulnerability descriptions, but are designed to support repair within a narrower scope than project-level artifacts. However, Table~\ref{tab:requirement_level_verified_outcome_summary} shows that complete report structure does not by itself lead to uniformly successful repair. In JavaScript, the reported guidance is sufficient for secure-and-correct outcomes in nearly all requirements, with only one case requiring bounded iteration. In Python, a similar pattern holds, although some requirements still need additional rounds before converging. In PHP, requirement-level repair is more mixed: some requirements succeed immediately, some converge only after several iterations, and others remain unsuccessful under verification. Taken together, Tables~\ref{tab:requirement_level_vd_guidance} and~\ref{tab:requirement_level_verified_outcome_summary} suggest that requirement-level VD reports can provide useful repair guidance, and in general are more actionable than project-level reports, but their real value still depends on whether the repair stage can turn that guidance into a patch that preserves functionality and removes exploitability.

\begin{table}
\centering
\caption{Requirement-level characteristics of detection guidance.}
\label{tab:requirement_level_vd_guidance}
\begin{tblr}{
  row{1} = {c},
  cell{2}{1} = {r=5}{},
  cell{2}{3} = {c},
  cell{2}{4} = {c},
  cell{2}{5} = {c},
  cell{2}{6} = {c},
  cell{2}{7} = {c},
  cell{3}{3} = {c},
  cell{3}{4} = {c},
  cell{3}{5} = {c},
  cell{3}{6} = {c},
  cell{3}{7} = {c},
  cell{4}{3} = {c},
  cell{4}{4} = {c},
  cell{4}{5} = {c},
  cell{4}{6} = {c},
  cell{4}{7} = {c},
  cell{5}{3} = {c},
  cell{5}{4} = {c},
  cell{5}{5} = {c},
  cell{5}{6} = {c},
  cell{5}{7} = {c},
  cell{6}{3} = {c},
  cell{6}{4} = {c},
  cell{6}{5} = {c},
  cell{6}{6} = {c},
  cell{6}{7} = {c},
  cell{7}{1} = {r=5}{c},
  cell{7}{3} = {c},
  cell{7}{4} = {c},
  cell{7}{5} = {c},
  cell{7}{6} = {c},
  cell{7}{7} = {c},
  cell{8}{3} = {c},
  cell{8}{4} = {c},
  cell{8}{5} = {c},
  cell{8}{6} = {c},
  cell{8}{7} = {c},
  cell{9}{3} = {c},
  cell{9}{4} = {c},
  cell{9}{5} = {c},
  cell{9}{6} = {c},
  cell{9}{7} = {c},
  cell{10}{3} = {c},
  cell{10}{4} = {c},
  cell{10}{5} = {c},
  cell{10}{6} = {c},
  cell{10}{7} = {c},
  cell{11}{3} = {c},
  cell{11}{4} = {c},
  cell{11}{5} = {c},
  cell{11}{6} = {c},
  cell{11}{7} = {c},
  cell{12}{1} = {r=5}{c},
  cell{12}{3} = {c},
  cell{12}{4} = {c},
  cell{12}{5} = {c},
  cell{12}{6} = {c},
  cell{12}{7} = {c},
  cell{13}{3} = {c},
  cell{13}{4} = {c},
  cell{13}{5} = {c},
  cell{13}{6} = {c},
  cell{13}{7} = {c},
  cell{14}{3} = {c},
  cell{14}{4} = {c},
  cell{14}{5} = {c},
  cell{14}{6} = {c},
  cell{14}{7} = {c},
  cell{15}{3} = {c},
  cell{15}{4} = {c},
  cell{15}{5} = {c},
  cell{15}{6} = {c},
  cell{15}{7} = {c},
  cell{16}{3} = {c},
  cell{16}{4} = {c},
  cell{16}{5} = {c},
  cell{16}{6} = {c},
  cell{16}{7} = {c},
  hlines = {dotted},
  hline{1-2,7,12,17} = {-}{solid,0.08em},
}
Language     & Requirement & {\#VD\\Reports} & {VD\\Complete} & {Avg.\\Entry\\Points} & {Avg.\\Evidence\\Files} & {Avg.\\Patch\\Locations} \\
JavaScript   & R1          & 5               & 100\%          & 3.00                  & 2.20                    & 2.60                     \\
             & R2          & 2               & 100\%          & 2.50                  & 3.00                    & 3.50                     \\
             & R3          & 2               & 100\%          & 4.00                  & 3.00                    & 2.00                     \\
             & R4          & 2               & 100\%          & 3.00                  & 2.50                    & 3.50                     \\
             & R5          & 2               & 100\%          & 3.50                  & 3.50                    & 3.50                     \\
PHP          & R1          & 5               & 100\%          & 4.00                  & 3.40                    & 5.20                     \\
             & R2          & 1               & 100\%          & 9.00                  & 6.00                    & 16.00                    \\
             & R3          & 2               & 100\%          & 4.50                  & 2.50                    & 3.00                     \\
             & R4          & 2               & 100\%          & 3.00                  & 2.00                    & 4.00                     \\
             & R5          & 2               & 100\%          & 3.50                  & 2.50                    & 4.00                     \\
Python       & R1          & 5               & 100\%          & 3.60                  & 3.00                    & 3.40                     \\
             & R2          & 2               & 100\%          & 2.50                  & 2.50                    & 3.50                     \\
             & R3          & 3               & 100\%          & 5.00                  & 3.00                    & 4.00                     \\
             & R4          & 4               & 100\%          & 4.25                  & 3.00                    & 5.50                     \\
             & R5          & 3               & 100\%          & 6.67                  & 3.67                    & 6.00                     \\
\end{tblr}
\end{table}

\begin{table}
\centering
\caption{Requirement-level verified repair outcomes under detection guidance.}
\label{tab:requirement_level_verified_outcome_summary}
\begin{tblr}{
  row{1} = {c},
  hlines,
  hline{3-4} = {-}{dotted},
}
Language   & Verified Workflow Outcome                                                                                  & Initial Interpretation                                                                                                                               \\
JavaScript & {Four of five requirements reach SC=1.0\\~in a single pass;\\R5 converges after one additional iteration.} & {Requirement-level guidance is consistently actionable,\\with bounded iteration needed only for the hardest case.}                                   \\
PHP        & {R1 and R2 succeed in a single pass;~\\R4 converges only at iteration 4;~\\R3 and R5 remain unsuccessful.} & {Requirement-level repair is mixed:~\\some requirements are tractable,~\\but others remain limited by persistent~\\regressions or incomplete fixes.} \\
Python     & {R2, R3, and R5 succeed in a single pass;~\\R1 converges at iteration 2;~\\R4 converges at iteration 4.}    & {Requirement-level guidance is generally useful, but\\trustworthy repair still depends on requirement complexity.}                                   
\end{tblr}
\end{table}

\begin{tcolorbox}[
  colback=gray!10,
  colframe=gray!60,
  boxrule=0.6pt,
  arc=2pt,
  left=6pt,right=6pt,top=4pt,bottom=4pt
]
\noindent \textbf{RQ2:} Detection reports are generally actionable for downstream repair because they usually provide explicit localization and repair guidance. Their reliability, however, is less uniform: structurally complete reports often align with successful repair, but in some settings the same level of guidance still does not lead to a secure-and-correct outcome under verification.
\end{tcolorbox}

\subsection{RQ3 (Repair trustworthiness and dominant failure modes).}

\subsubsection{Measures}
RQ3 evaluates repair trustworthiness using verified repair outcomes. A repair is considered trustworthy when it reaches a secure-and-correct outcome, meaning that the reported vulnerability is no longer exploitable and the repaired artifact still passes the corresponding functional tests. Repairs that do not satisfy both conditions are treated as untrustworthy.

To compare trustworthiness across scopes, the results are summarized separately for the project-level, requirement-level, and file-level settings. For each setting, the analysis distinguishes between repairs that succeed in a single pass, repairs that converge only after bounded iteration, and repairs that do not reach a secure-and-correct state within the repair budget. The same verified outcomes are also used to describe the main failure patterns, including cases that remain exploitable, cases that lose functional correctness, and cases that require additional iterations before both conditions are restored.
\subsubsection{Findings.}
\paragraph{Project-level observations.}
Table~\ref{tab:rq3_project_level_trustworthiness} shows that project-level repair is the least stable setting among the evaluated scopes. JavaScript reaches a secure-and-correct outcome in a single pass, indicating that the repair remains trustworthy under verification. Python also reaches a secure-and-correct state, but only after bounded iteration, which means that the early repair attempts are not yet reliable. PHP does not converge to a secure-and-correct outcome within the allowed budget, even after repeated repair attempts. Overall, these results suggest that project-level repair can succeed, but its trustworthiness is uneven and depends strongly on whether iteration can recover both security and functionality.

\begin{table}
\centering
\small
\caption{Project-level repair trustworthiness under verification.}
\label{tab:rq3_project_level_trustworthiness}
\begin{tblr}{
  width = \linewidth,
  colspec = {Q[l,1.2cm] X[1.8,l] X[1.5,l] X[2.1,l]},
  hlines,
  hline{3-4} = {-}{dotted},
}
Language   & Verified Outcome Summary & Dominant Failure Pattern & Initial Interpretation \\
JavaScript & Secure-and-correct in a single pass. & No dominant failure observed in the final verified result. & Project-level repair is trustworthy in this setting, as the reported guidance aligns with a verified SC outcome. \\
PHP        & Single-pass repair fails, and iterative DRV does not converge to SC within four iterations. & Persistent regression and incomplete repair. & Project-level repair is not trustworthy in this setting, because repeated repair attempts do not restore both functionality and security. \\
Python     & Single-pass repair fails, but iterative DRV reaches SC at iteration 3. & Delayed convergence, with earlier iterations not yet trustworthy. & Project-level repair can become trustworthy, but only after bounded iteration restores both exploit resistance and functional correctness. \\
\end{tblr}
\end{table}

\paragraph{Requirement-level observations.}
Table~\ref{tab:rq3_requirement_level_trustworthiness} shows that requirement-level repair is more stable than project-level repair, although the outcome still depends on the target requirement. In JavaScript, four of the five requirements reach secure-and-correct outcomes in a single pass, and the remaining case converges after one additional iteration. Python follows a similar pattern, but two requirements require bounded iteration before both security and functionality are restored. PHP is less consistent: R1 and R2 succeed immediately, R4 converges only at iteration 4, and R3 and R5 remain unsuccessful under verification. Overall, narrowing the repair scope improves repair trustworthiness, but difficult requirements can still fail to preserve both exploit resistance and intended behavior.

\begin{table}
\centering
\small
\caption{Requirement-level repair trustworthiness under verification.}
\label{tab:rq3_requirement_level_trustworthiness}
\begin{tblr}{
  width = \linewidth,
  colspec = {Q[l,1.2cm] X[1.8,l] X[1.5,l] X[2.1,l]},
  hlines,
  hline{3-4} = {-}{dotted},
}
Language   & Verified Outcome Summary & Dominant Failure Pattern & Initial Interpretation \\
JavaScript & Four of five requirements are secure-and-correct in a single pass; the remaining case (R5) converges at iteration 2. & Isolated incomplete repair in the hardest requirement. & Requirement-level repair is highly trustworthy, with bounded iteration sufficient for the only hard case. \\
PHP        & R1 and R2 succeed in a single pass; R4 converges at iteration 4; R3 and R5 remain unsuccessful. & Persistent regression and incomplete repair in selected requirements. & Requirement-level trustworthiness is mixed: some requirements are repairable, while others fail to preserve functionality or fully remove exploitability. \\
Python     & R2, R3, and R5 succeed in a single pass; R1 converges at iteration 2; R4 converges at iteration 4. & Delayed convergence in requirements with broader repair scope. & Requirement-level repair is generally trustworthy, but some requirements need additional iterations before both security and functionality are restored. \\
\end{tblr}
\end{table}

\paragraph{File-level observations.}
Table~\ref{tab:rq3_file_level_trustworthiness} shows that file-level repair is the most stable setting under verification. For both ChatGPT-5 and GLM-5, most files reach secure-and-correct outcomes in a single pass, and the few remaining cases converge after bounded iteration. Compared with project-level and requirement-level repair, failures at this level are more limited in scope and are less likely to involve broad functional disruption. The remaining unsuccessful cases are concentrated in a small number of difficult files, where repair either misses part of the vulnerable logic or requires additional rounds before both security and functionality are restored. Overall, restricting repair to a single vulnerable file yields the most trustworthy outcomes across the evaluated settings.

\begin{table}
\centering
\small
\caption{File-level repair trustworthiness by model under verification.}
\label{tab:rq3_file_level_trustworthiness}
\begin{tblr}{
  width = \linewidth,
  colspec = {Q[l,1.6cm] X[2.0,l] X[1.4,l] X[1.8,l]},
  hlines,
  hline{3} = {-}{dotted},
}
Model & Verified Outcome Summary & Dominant Failure Pattern & Initial Interpretation \\
ChatGPT-5 & Most files reach secure-and-correct outcomes in a single pass; only a few difficult files require bounded iteration. & Localized incomplete repair or isolated regression. & File-level repair is highly trustworthy, with failures confined to a very small number of difficult vulnerabilities. \\
GLM-5 & Most files also reach secure-and-correct outcomes in a single pass; only a few hard cases require iterative refinement. & Delayed convergence in localized hard cases. & File-level repair is also highly trustworthy, and remaining difficulty is limited to a small number of files. \\
\end{tblr}
\end{table}

\paragraph{Failure patterns under verification.}
The verification results show two dominant failure modes. The first is incomplete repair, in which the vulnerability remains exploitable or is only partially mitigated after patching. The second is regression-related failure, in which the repair disrupts previously correct behavior and therefore fails to reach a secure-and-correct state. These two patterns account for most unsuccessful cases at the project and requirement levels. By contrast, file-level failures are fewer and are usually confined to delayed convergence in a small number of difficult cases. Overall, the results indicate that trustworthy repair is limited less by the absence of repair attempts than by the difficulty of combining vulnerability mitigation with behavioral preservation.

\begin{tcolorbox}[
  colback=gray!10,
  colframe=gray!60,
  boxrule=0.6pt,
  arc=2pt,
  left=6pt,right=6pt,top=4pt,bottom=4pt
]
\noindent \textbf{RQ3:} LLM-based repair can produce trustworthy fixes, but the result depends strongly on repair scope. Trustworthiness is lowest at the project level, improves at the requirement level, and is highest at the file level. Under verification, the main remaining difficulty is that some cases do not reach a secure-and-correct state within the bounded budget, while others require additional iterations before both security and functionality are restored.
\end{tcolorbox}

\section{Discussion}
\label{disscusion}

This section discusses the main implications of RQ1--RQ3 for LLM-based vulnerability management.

\subsection{Pipeline-level behavior under iteration}
RQ1 shows that bounded iterative DRV can improve secure-and-correct yield, but the improvement is not uniform. The clearest gains appear when the repair target is narrower. At the file level, the model is more often able to make a local change that removes the targeted vulnerability without disturbing unrelated behavior. At the project level, the same loop is less stable. Broader context may help the model reason about the system, but it also increases ambiguity about where the real cause lies and how widely a repair should spread.

This helps explain why iteration does not automatically lead to better outcomes. Additional rounds are useful only when each round produces a change that can be checked clearly by verification. When detection guidance is specific and the edit stays local, later iterations can refine the repair and recover from earlier mistakes. When guidance is vague, or when the repair touches multiple files and interfaces, later iterations often spend effort dealing with side effects rather than removing the original vulnerability. In that case, iteration becomes a way of managing instability rather than making steady progress.

\subsection{Detection as repair guidance}
RQ2 shows that detection reports are often useful for downstream repair, but usefulness alone is not enough. Many reports provide concrete information such as an affected endpoint, a relevant parameter, or a likely access-control boundary. That kind of detail helps the repair step focus on a plausible location and makes the next verification round more informative. In this sense, many reports are actionable.

At the same time, the results show that actionable does not always mean reliable. A report may look detailed but still describe the wrong cause, point to the wrong place, or oversimplify how the vulnerability is triggered in the actual workflow. When that happens, the repair may still produce code changes, but those changes do not reliably lead to a secure-and-correct outcome. This problem becomes more serious as repair scope grows. At narrower scopes, a partially correct report may still be enough because the number of plausible edits is small. At broader scopes, the same report can leave too much uncertainty, and the model is more likely to patch symptoms rather than the underlying cause.

\subsection{Repair trustworthiness under verification}
RQ3 shows that repair trustworthiness depends strongly on verification and on repair scope. A repair should not be considered successful only because the code looks reasonable or because the model claims to have fixed the issue. It is trustworthy only when the targeted exploit is no longer reproducible and the intended functionality still works. Both checks matter. Security fixes often change validation logic, authorization checks, or control flow, and those are exactly the places where functional behavior can be altered by accident.

The observed outcomes fall into a few recurring cases. In the best case, the vulnerability is mitigated and functionality is preserved. In other cases, the security signal improves but the repair introduces a regression, often by restricting behavior too aggressively or by changing an interface that other parts of the system still depend on. There are also cases where the patch looks plausible but does not change the verified result: the vulnerable path remains reachable, the fix is incomplete, or the repair is applied to the wrong place. Under bounded iteration, some of these cases improve after additional rounds, but others remain unresolved within the available budget. This is why verification should be treated as the basis for trust, not as a final formality.

\subsection{Implications for practice and benchmarking}
For practice, the results support a simple lesson: verification should stay inside the repair loop. Running both exploit tests and functional tests after each repair attempt makes the process evidence-driven. It also helps distinguish real progress from changes that only appear plausible. The results further suggest that repair scope should be chosen with care. Broader-scope repair gives the model more context, but it also increases the chance of ambiguous localization and wider side effects. Narrower-scope repair is often easier to verify and more likely to produce stable outcomes.

For benchmarking, the results show why runnable, test-grounded, multi-granularity evaluation matters. A project-only benchmark would show that broad repair is difficult, but it would miss how much more stable repair can become when the target is narrowed. A file-only benchmark would show stronger repair performance, but it would understate the coordination problems that appear at larger scales. By spanning project-level, requirement-level, and file-level settings, the benchmark makes these differences visible. More generally, runnable artifacts paired with executable functional and exploit tests make it possible to observe not only whether a vulnerability disappears, but also whether the repair preserves the behavior that the software is supposed to provide.

\section{Threats to Validity}
\label{threats}

\subsection{Internal Validity}

\textbf{Test-grounded outcomes and oracle coverage.}
Secure-and-correct yield is determined by executable verification: functional tests are used to assess correctness, while exploit tests and post-repair re-detection are used to assess security. The results therefore depend on the coverage and adequacy of these oracles. To reduce this threat, the same artifacts, test suites, and verification procedure are used consistently across workflow variants and across repair iterations.

\textbf{Bounded iteration and model variance.}
The study uses bounded iteration budgets to keep comparisons fair across settings, but some unresolved cases may reflect the chosen budget rather than the absolute inability of the workflow to recover. In addition, LLM outputs and model-based detection results can vary across runs. To limit this effect, the study uses matched workflow settings, stopping rules, and verification procedures across variants, while interpreting the remaining variation as part of practical workflow behavior.

\subsection{External Validity}

\textbf{Scope of the benchmark.}
The benchmark is built from runnable web-application artifacts in PHP, JavaScript, and Python. This makes the study suitable for web-oriented vulnerability management, but the findings should not be assumed to transfer directly to other domains such as systems software, mobile applications, embedded software, or smart contracts.

\textbf{Granularity and vulnerability coverage.}
The study covers project-level, requirement-level, and file-level settings, but it does not represent all possible interaction styles for LLM-assisted development and repair. In addition, the benchmark includes targeted vulnerability instances and matching exploit tests rather than the full space of real-world vulnerability classes. Broader workflows, richer application features, and additional vulnerability types may lead to different results.

\section{Conclusion}
\label{conclusion}

This study examined LLM-based vulnerability management through a bounded Detect--Repair--Verify (DRV) workflow and supported the evaluation with EduCollab, a multi-language benchmark of runnable LLM-generated web applications paired with executable functional and exploit tests. The results show that iterative DRV can improve secure-and-correct yield, but the benefit depends strongly on repair scope: broader scopes are less stable and more likely to introduce side effects, whereas narrower scopes are easier to verify and more likely to produce secure-and-correct outcomes. The study also finds that detection reports can assist downstream repair, but detailed reports do not necessarily provide reliable guidance, and repair quality cannot be judged from patches alone without verification of both vulnerability removal and functional preservation. Overall, the findings suggest that LLM-based vulnerability management should be treated as a test-grounded iterative process rather than a one-shot repair task, and that evaluation should be performed on runnable artifacts under executable functional and security tests.

\section*{Declarations}

\subsection*{Funding}
This research did not receive any external funding.

\subsection*{Ethical approval}
Not applicable.

\subsection*{Informed consent}
Not applicable.

\subsection*{Author Contributions}
Cheng Cheng: Conceptualization, methodology, data collection, writing---original draft, and writing---review and editing.\\

\subsection*{Data Availability Statement}
The replication package is available at: \texttt{https://github.com/Hahappyppy2024/EmpricalVDR}.

\subsection*{Conflict of Interest}
The author declare that they have no competing interests.

\subsection*{Clinical trial number}
Clinical trial number: not applicable.

\bibliographystyle{spbasic}      

\bibliography{ref}   

\end{document}